\RequirePackage{pdf14}

\pdfoutput=1
\documentclass[letterpaper]{article}
\usepackage[usenames,dvipsnames,table]{xcolor}
\usepackage{graphicx}
\usepackage{caption}
\usepackage{subcaption}
\usepackage{array,setspace,mathrsfs,amsthm,amsfonts,amsmath,yfonts,dsfont,bbm,colonequals,mathtools}
\usepackage{./aux/jheppub}
\usepackage{afterpage}
\usepackage{xspace}
\usepackage[normalem]{ulem}

\def\d{\mathrm{d}}

\def\zb{{\bar{z}}}

\def\log{\mathrm{log}}

\def\AA{{\mathcal{A}}}
\def\BB{{\mathcal{B}}}
\def\CC{{\mathcal{C}}}

\def\EE{{\mathcal{E}}}
\def\OO{{\mathcal{O}}}
\def\NN{{\mathcal{N}}}

\def\GG{{\mathcal{G}}}
\def\FF{{\mathcal{F}}}

\def\MM{{\mathcal{M}}}

\def\dDisc{\mathrm{dDisc}}

\def\SU{\mathrm{SU}}
\def\U{\mathrm{U}}

\newcommand{\be}{\begin{equation}}
\newcommand{\ee}{\end{equation}}
\newcommand{\bea}{\begin{eqnarray}}
\newcommand{\eea}{\end{eqnarray}}
\newcommand{\nn}{\nonumber}

\newcommand\eg{{\it e.g.}}
\newcommand\ie{{\it i.e.}}

\title{Bootstrapping the $(A_1,A_2)$ Argyres-Douglas theory}
\author{Martina Cornagliotto,}
\author{Madalena Lemos,}
\author{Pedro Liendo}

\affiliation{DESY Hamburg, Theory Group, Notkestra{\ss}e 85, D-22607 Hamburg, Germany}

\emailAdd{$\lbrace$martina.cornagliotto,madalena.lemos,pedro.liendo$\rbrace$@desy.de}

\preprint{DESY 17-175}

\bigskip
\abstract{We apply bootstrap techniques in order to constrain the CFT data of the  $(A_1,A_2)$ Argyres-Douglas theory, which is arguably the simplest of the Argyres-Douglas models. We study the four-point function of its single Coulomb branch chiral ring generator and put numerical bounds on the low-lying spectrum of the theory. Of particular interest is an infinite family of semi-short multiplets labeled by the spin $\ell$. Although the conformal dimensions of these multiplets are protected, their three-point functions are not. Using the numerical bootstrap we impose rigorous upper and lower bounds on their values for spins up to $\ell=20$.
Through a recently obtained inversion formula, we also estimate them for sufficiently large $\ell$, and the comparison of both approaches shows consistent results. 
We also give a rigorous numerical range for the OPE coefficient of the next operator in the chiral ring, and estimates for the dimension of the first R-symmetry neutral non-protected multiplet for small spin.}


\begin{document}
\setcounter{tocdepth}{2}
\maketitle
\setcounter{page}{1}

\section{Introduction and summary}
\label{sec:intro}

The revival of the conformal bootstrap program \cite{Rattazzi:2008pe}, has provided new tools to study non-perturbative physics. The numerical techniques introduced in \cite{Rattazzi:2008pe} have given a wealth of results, with the most impressive being the high-precision estimates of the critical exponents of the $3d$ Ising model \cite{ElShowk:2012ht,El-Showk:2014dwa,Kos:2014bka,Simmons-Duffin:2015qma,Kos:2016ysd}. 
In a parallel line of development, analytic approaches to the bootstrap have also been explored, and recent progress has given access to the spectrum of conformal field theories (CFTs) at large spin by means of the lightcone limit \cite{Fitzpatrick:2012yx,Komargodski:2012ek}.
These two methods were combined in \cite{Alday:2015ota,Simmons-Duffin:2016wlq}, where knowledge of
operator dimensions and operator product expansion (OPE) coefficients, obtained numerically for the Ising model, was used to derive analytic approximations for the CFT data at large spin. Remarkably, the analytic results obtained matched the numerical data down to spin two.

The \emph{superconformal} bootstrap has also seen substantial progress. Apart from numerical explorations of crossing symmetry \cite{Poland:2010wg,Poland:2011ey,Poland:2015mta,Li:2017ddj,Berkooz:2014yda,
Beem:2014zpa,Beem:2013qxa,Alday:2013opa,Alday:2014qfa,
Chester:2014fya,Chester:2014mea,Lemos:2015awa,Chester:2015qca,
Bobev:2015vsa,Bobev:2015jxa,Bae:2016jpi,Beem:2015aoa,Lin:2015wcg,Lin:2016gcl,Beem:2016wfs,Lemos:2016xke,
Cornagliotto:2017dup,Chang:2017xmr,Chang:2017cdx}, the bootstrap line of thinking helped uncover a solvable subsector in four-dimensional superconformal theories \cite{Beem:2013sza}.\footnote{See also \cite{Beem:2014kka} and \cite{Chester:2014mea,Beem:2016cbd} for similar results in six and three dimensions.} More precisely, the results of \cite{Beem:2013sza} imply that any $4d$ $\NN \geqslant 2$ superconformal field theory (SCFT) contains a closed subsector isomorphic to a $2d$ chiral algebra.

The subsector captures certain protected quantities, and in order to access non-protected data the numerical bootstrap is still a necessary tool. In some cases, similarly to the $3d$ Ising model and $O(N)$ models, known supersymmetric theories appear on special points, ``kinks'', of the numerically produced exclusion curves, and the numerical machinery of \cite{Rattazzi:2008pe} can be applied in order to extract the CFT data.
However, kinks seem to be scarce, in particular, while the numerical bounds for $4d$ $\NN=2$ SCFTs obtained in \cite{Beem:2014zpa,Lemos:2015awa} put strong constraints on the landscape of theories, they did not single out any particular solution to crossing. 

\medskip

In this work we focus on the ``simplest'' four-dimensional $\NN=2$ Argyres-Douglas SCFT: the $(A_1,A_2)$ (or $H_0$) theory \cite{Argyres:1995jj,Argyres:1995xn}. It has the lowest possible $c$-anomaly coefficient among interacting SCFTs \cite{Liendo:2015ofa}, and the lowest $a$-anomaly coefficient among the known ones. The $(A_1,A_2)$ SCFT can be realized by going to a special point on the Coulomb branch of an $\NN=2$ supersymmetric gauge theory, with gauge group $\SU(3)$, where electric and magnetic particles become simultaneously massless \cite{Argyres:1995jj,Argyres:1995xn}.
It is an isolated $\NN=2$ SCFT, with no exactly marginal deformations, and thus no weak-coupling description. As such, despite being known for a very long time, little is known about the spectrum of this theory.
Known data includes the scaling dimension, $\Delta_{\phi}$, of the single generator of the Coulomb branch chiral ring, whose vev parametrizes the Coulomb branch, and the $a$- and $c$-anomaly coefficients \cite{Aharony:2016kai}:
\be 
\Delta_{\phi} = \frac{6}{5}\, , \qquad c = \frac{11}{30}\, , \qquad a = \frac{43}{120}\, .
\label{eq:acanomalies_dimension}
\ee
The full superconformal index \cite{Kinney:2005ej,Gadde:2011uv,Rastelli:2014jja} was recently computed using an $\NN=1$ Lagrangian that flows to the $(A_1,A_2)$ SCFT in the IR \cite{Maruyoshi:2016tqk}. 
The chiral algebra of this theory is conjectured to be the Yang-Lee minimal  model \cite{rastelli_harvard,Beem:2017ooy}, which gives access to the spectrum of a particular class of short operators, dubbed ``Schur'' operators.
However, the chiral algebra is insensitive to the Coulomb branch data of the theory, and even though the dimensions of the  operators parameterizing the Coulomb branch chiral ring are known, not much is known about the values of the corresponding three-point functions.\footnote{See \cite{Hellerman:2017sur} for a recent computation of the two-point function (in normalizations where the OPE coefficients are one) of a Coulomb branch chiral ring operator, for theories with a single chiral ring generator, in the limit of large $\U(1)_r$ charge.}

The relatively low values of its central charge and of the dimension of its Coulomb branch chiral ring generator make the $(A_1,A_2)$ Argyres-Douglas theory amenable to numerical bootstrap techniques. 
In fact, one could argue that this is the $\NN=2$ SCFT with the best chance to be ``solved'' numerically.
We approach this theory based on the existing Coulomb branch data, by considering four-point functions of $\NN=2$ chiral operators, whose superconformal primaries are identified with the elements of the Coulomb branch chiral ring.\footnote{Another natural operator to consider in the correlation functions would be the $\NN=2$ stress-tensor multiplet, however, the superconformal blocks for this multiplet are not known, and we leave this for future work.}
While the values of $c$ and $\Delta_\phi$ in \eqref{eq:acanomalies_dimension} are not selected by the numerical bootstrap, thanks to supersymmetry they are exactly known and thus we can use them as input in our analysis.
We note however, that nothing is known about the spectrum of 
non-supersymmetry preserving
relevant deformations of the $(A_1,A_2)$ theory, and this type of information was essential to corner the $3d$ Ising model to a small ``island'' \cite{Kos:2014bka}. 

\bigskip

The results we find are encouraging, and provide the first estimates for unprotected quantities in this theory.
We start by obtaining a lower bound on the central charge valid for any $\NN=2$ theory with a Coulomb branch chiral ring operator of dimension $\Delta_{\phi} = \frac65$. This bound appears to be converging to a value \emph{close} to $c=\tfrac{11}{30}$, however the numerics are not conclusive enough. If the bound on $c$ converges to $\tfrac{11}{30}$, then there is a unique solution to the crossing equations at $\Delta_{\phi} = \frac65$ that corresponds to the $(A_1,A_2)$ theory.
If the numerical bound falls short of $\tfrac{11}{30}$, we present evidence, in the form of valid bounds on OPE coefficients and estimates on operator dimensions, that the various solutions around $c \sim \tfrac{11}{30}$ do not look so different, as far as certain observables are concerned. 
While the results we obtain are not at the level of the precision numerics of the $3d$ Ising model, 
we are able to provide estimates for the CFT data of this theory.
For example, we constrain the OPE coefficient of 
the square of the Coulomb branch chiral ring generator (after unit normalizing its two-point function) to lie in the interval
\be
2.1418 \leqslant \lambda_{\EE_{\frac{12}5}}^2 \leqslant 2.1672 \,.
\label{eq:Ebound_intro}
\ee
While this is a true bound, due to slow convergence it is still far from being optimal, and will improve as more of the constraints of the crossing equations are taken into account. In section \ref{sec:OPEbounds} we present estimates for the optimal range, based on conservative extrapolations of the bounds.
Similarly, we constrain the OPE coefficients of a family of semi-short multiplets,
 appearing in the self-OPE of $\NN=2$ chiral operators, to lie in a narrow range, quoted in \eqref{eq:Clbound} for $\ell=2,4$, and in figure \ref{Fig:ClCH} for even spins up to $\ell=20$.

We also provide in \eqref{eq:dimextrapol} the first estimate of the dimension of the lowest-lying unprotected scalar appearing in the  OPE of the $\NN=2$ chiral operator with its conjugate. This operator corresponds to a long multiplet that is a singlet under $\SU(2)_R$ symmetry, and neutral under $U(1)_r$, and we find it is relevant.
These estimates are obtained from the extremal functionals \cite{ElShowk:2012hu} that gave rise to the aforementioned OPE coefficient bounds. From these extremal functionals we also obtain rough estimates for the dimensions of the lowest-twist long operator for higher values of the spin, shown in figure \ref{Fig:anomCH}. Surprisingly, for spin greater than zero these operators are very close to being double-twist operators, \ie, $\Delta = 2 \Delta_\phi + \ell$.

Finally, we make use of the inversion formula of \cite{Caron-Huot:2017vep} to obtain large-spin estimates of the CFT data. As our numerical results are much further away from convergence than \cite{Simmons-Duffin:2016wlq}, we refrain from using them as input in the inversion formula. As such the only input we provide is the identity and stress-tensor supermultiplet exchange (with the appropriate central charge). Interestingly, we find that this input already provides a reasonable estimate of the numerically-bounded quantities for small spin. 

A hybrid approach, combining both the numerical bootstrap and the inversion formula seems to be the most promising way to proceed, perhaps along the lines of the one suggested in \cite{Simmons-Duffin:2016wlq}. The results of this paper are a first step in this direction, and give us hope that a large amount of CFT data can be bootstrapped for the $(A_1,A_2)$ theory.
\section{The \texorpdfstring{$(A_1,A_2)$}{(A1,A2)} Argyres-Douglas theory}
\label{sec:h0background}

Argyres-Douglas theories \cite{Argyres:1995jj,Argyres:1995xn} were first obtained by going to a special point on the Coulomb branch of an $\NN=2$ theory in which several BPS particles, with mutually non-local charges, become massless simultaneously. 
Among the various Argyres-Douglas models a particular class appears to be the ``simplest'', that is, the $(A_1,A_{2n})$ theories obtained in \cite{Eguchi:1996vu}. 
They are rank $n$ theories, 
\ie, their Coulomb branches have complex dimension $n$, and have trivial Higgs branches \cite{Argyres:2012fu}. 
The  chiral algebras associated to the $(A_1,A_{2n})$ theories have been conjectured to be the non-unitary series of Virasoro minimal models $\MM_{2,2n+3}$ \cite{rastelli_harvard,Beem:2017ooy}, and in this sense the theories could be argued to be ``simple''. 

In this paper we focus on the $n=1$ case of the $(A_1,A_{2n})$ Argyres-Douglas family, which is of rank one and thus the simplest in this class. In fact, among all interacting rank one SCFTs obtained through the systematic classification of \cite{Argyres:2015ffa,Argyres:2015gha,Argyres:2016xua,Argyres:2016xmc,Argyres:2016yzz}, it corresponds to the theory with the smallest $a$-anomaly coefficient,\footnote{This assumes the standard lore that the Coulomb branch chiral ring is freely generated. While this is true for all known SCFTs, there is no proof of this fact in a generic SCFT. See \cite{Argyres:2017tmj} for an exploration of theories which could have relations on the Coulomb branch.} which provides a measure of degrees of freedom in CFT \cite{Komargodski:2011vj}.
This theory was originally obtained on the Coulomb branch of a pure $\SU(3)$ gauge theory, or alternatively from an $\SU(2)$ gauge theory with a single hypermultiplet \cite{Argyres:1995jj,Argyres:1995xn}.
There is no standard nomenclature for this model, and in this paper we follow the $(A_1,A_2)$ naming convention based on its BPS quiver \cite{Cecotti:2010fi}. To emphasize its original construction it was also named $AD_{N_f=0}(\SU(3))$ and $AD_{N_f=1}(\SU(2))$  in \cite{Tachikawa:2013kta}. Finally, it can also be realized in  F-theory, on a single D3-brane probing a codimension one singularity of type $H_0$ where the dilaton is constant \cite{Dasgupta:1996ij,Aharony:1998xz}. 
For this reason the theory is often referred to as the $H_0$ theory.

The $(A_1,A_2)$ theory is an intrinsically interacting isolated fixed point with no marginal coupling: it does not have a conformal manifold nor a weak-coupling expansion. Recently, there has been progress in obtaining RG flows from $\NN=1$ Lagrangian theories that end on Argyres-Douglas SCFTs in the IR \cite{Maruyoshi:2016aim,Maruyoshi:2016tqk,Agarwal:2016pjo,Agarwal:2017roi,Benvenuti:2017bpg}, and in particular the $(A_1,A_2)$ theory can be obtained starting from a deformation of $\SU(2)$  $\NN=2$ superconformal QCD. This allows for the computation of some information about the theory, such as the superconformal index. 
As quoted in \eqref{eq:acanomalies_dimension} the values of the $a$- and $c$-anomaly coefficients are known, first obtained through a holographic computation in \cite{Aharony:2016kai},
and the dimension of the single generator of the Coulomb branch chiral ring is also known, and given in~\eqref{eq:acanomalies_dimension}.
Coulomb branch chiral ring operators can be associated with the scalar primaries of $\NN=2$ chiral operators, $\EE_{r}$ multiplets in the notation of \cite{Dolan:2002zh}, and this implies this SCFT must contain an operator  with $r_0=\Delta_{\phi}=\frac65$, as well as its conjugate.

The chiral algebra of the $(A_1,A_2)$ theory is conjectured to be the $\MM_{2,5}$ minimal model \cite{rastelli_harvard,Beem:2017ooy}, also known as the Yang-Lee edge singularity.
The first indication of this conjecture comes from the central charge. The basic chiral algebra dictionary states that $4d$ and $2d$ central charges are related by $c_{2d} = -12c_{4d}$; for the $(A_1,A_2)$ theory this gives $c_{2d} = -\frac{22}{5}$, which is indeed the correct value for the Yang-Lee model. 
Thanks to the interplay between $2d$ and $4d$ descriptions one can actually prove that $c_{4d} \geqslant \frac{11}{30}$ for any interacting $\NN=2$ SCFTs \cite{Liendo:2015ofa}.\footnote{Similar bounds can be obtained for $\NN=3$ \cite{Cornagliotto:2017dup} and $\NN=4$ \cite{Beem:2013qxa,Beem:2016wfs} theories, and also for $\NN=2$ theories with flavor symmetries  \cite{Beem:2013sza,Lemos:2015orc,Beem:2017ooy}.} This bound is saturated by the  $(A_1,A_2)$ theory, which in some sense sits at the origin of the $\NN=2$ theory space, as all other interacting SCFTs must have higher values of the $c$-central charge.
Another entry of the chiral algebra dictionary states that the Schur limit of the superconformal index \cite{Kinney:2005ej,Gadde:2011uv,Rastelli:2014jja} should match the $2d$ vacuum character. For the Yang-Lee minimal model the vacuum character seems to match the expression for the Schur index proposed in \cite{Cordova:2015nma}, while the character of the non-vacuum module has been matched to the index in the presence of a surface defect \cite{Cordova:2017mhb}.
Using the Yang-Lee model we can compute three-point functions of Schur operators, \ie, the operators captured by the chiral algebra,
modulo ambiguities when lifting operators from the $2d$ chiral algebra to representations of the four-dimensional superconformal algebra. A conjectured prescription on how to lift these ambiguities for the $(A_1,A_2)$ Argyres-Douglas theory has been put forward in \cite{Song:2016yfd}. Coulomb branch chiral ring operators, however, are not captured by the chiral algebra.

The features described above suggest that the $(A_1,A_2)$ theory might be the simplest $\NN=2$ interacting SCFT. Despite this, apart from the aforementioned quantities not much is known about the CFT data of this theory.
The fact that the theory has a Coulomb branch operator of relatively low dimension, $r_0=\frac65$, and a very low $c$ central charge, makes it well suited for the bootstrap program. 
Hence, the goal of this paper is to use modern bootstrap tools in order to access \emph{non-protected} dynamical data. 
The natural first step
is to study the two operators that are guaranteed to be present:
the stress tensor, and the $\NN=2$ chiral operator that parametrizes the Coulomb branch. Since the superconformal blocks of the former remain elusive we focus on the latter.
A preliminary analysis of chiral correlators was already started in \cite{Beem:2014zpa,Lemos:2015awa}, however the main goal of those papers was the exploration of the landscape of $\NN=2$ SCFTs through their Coulomb branch data. 
In the following sections we instead focus exclusively on the $(A_1,A_2)$ theory, and attempt to ``zoom in'' on it by studying an $\NN=2$ chiral operator of fixed dimension $r_0=\frac65$.\footnote{There are other known SCFTs with a Coulomb branch chiral ring operator of dimension $r_0=\frac65$, in particular higher rank theories whose lowest dimensional Coulomb branch generator has this dimension are obtained in F-theory by probing a singularity of type $H_0$ with $N$ D3-branes \cite{Dasgupta:1996ij,Aharony:1998xz}. However, these theories have larger values of the $c$-anomaly coefficient \cite{Aharony:2016kai}, and by fixing the central charge we can focus on the $(A_1,A_2)$ theory.}


\subsection{OPE decomposition and crossing symmetry}

As we just discussed, our angle to attack the $(A_1,A_2)$ theory is through its Coulomb branch, and thus we are interested in the  $\NN=2$ chiral and anti-chiral operators, respectively $\EE_{r}$ and $\bar{\EE}_{r}$ multiplets, using the naming conventions of \cite{Dolan:2002zh}. 
These are short representations of the superconformal algebra that are half-BPS, where the superconformal primary  is annihilated by all supercharges of one chirality. We denote the superconformal primary of the chiral (anti-chiral) multiplets $\EE_{r}$  ($\bar{\EE}_{r}$) by  $\phi_{r}$ ($\bar{\phi}_{-r}$), where $r$ is the $U(1)_r$ charge of the superconformal primary, with unitarity requiring $r \geqslant 1$ ($-r \geqslant 1$).
The dimensions of the superconformal primaries $\phi_{r}$ ($\bar{\phi}_{-r}$) are fixed in terms of their $U(1)_r$ charges by $\Delta_\phi= r$ ($\Delta_{\bar{\phi}}=-r$). We refer the reader to, \eg, \cite{Dolan:2002zh}, for more on representation theory of the $\NN=2$ superconformal algebra.

The numerical bootstrap program applied to chiral correlators was considered in \cite{Beem:2014zpa,Lemos:2015awa} for the case of two identical operators, and their conjugates, and in \cite{Lemos:2015awa} for two distinct operators, and their conjugates.
Here we briefly review the setup for two identical operators $\EE_{r}$, and conjugates, and refer the reader to  \cite{Beem:2014zpa,Lemos:2015awa} for a more detailed account.
Considering all four-point functions involving the superconformal primaries of these multiplets, we write down the OPE selection rules and conformal block decompositions for all of the channels, and the crossing equations to be studied in sections \ref{sec:numericsh0} and \ref{sec:anlytical}. 
In this work we are only concerned with the $(A_1,A_2)$ theory and thus we fix $r$ to $r_0=\frac65$, according to \eqref{eq:acanomalies_dimension}.

\subsubsection{Non-chiral channel}
The OPE selection rules in the non-chiral channel are \cite{Beem:2014zpa}
\be
\phi_{r} \times \bar{\phi}_{-r}  \sim \mathbf{1} + \hat{\CC}_{0 (j,j)} + \AA^{\Delta > 2j+2}_{0, 0 (j,j)}\, .
\label{eq:selrulesnonchiral}
\ee
Here the $\hat{\CC}_{0 (j,j)}$ multiplets include conserved currents of spin $2j+2$, which for $j >0$ are absent in interacting theories \cite{Maldacena:2011jn,Alba:2013yda} and thus we will set them to zero. The multiplet $\hat{\CC}_{0(0,0)}$ corresponds to the superconformal multiplet that contains the stress tensor. By an abuse of notation we will often replace the subscript $(j,j)$ by $\ell$, with $\ell=2j$.
The superconformal block decomposition in this channel can be written as
\be
\langle \phi_{r}(x_1) \bar{\phi}_{-r}(x_2) \phi_{r}(x_3) \bar{\phi}_{-r}(x_4)  \rangle = 
\frac{1}{x_{12}^{2\Delta_{\phi}}x_{34}^{2\Delta_{\phi}}} \sum_{\OO_{\Delta,\ell}} |\lambda_{\phi \bar{\phi} \OO}|^2 \GG_{\Delta, \ell}(z,\bar{z})\, ,
\label{eq:nonchiral}
\ee
where the superblocks $\GG_{\Delta, \ell}(z,\bar{z})$, capturing the supersymmetric multiplets being exchanged in \eqref{eq:selrulesnonchiral}, were computed in \cite{Fitzpatrick:2014oza},
\be 
\label{eq:superblock}
\GG_{\Delta, \ell}(z,\bar{z}) = (z \bar{z})^{-\frac{\NN}{2}}g_{\Delta + \NN,\ell}^{\NN,\NN}(z,\bar{z})\, .
\ee
Here we wrote the blocks for $\NN=1,2$ chiral operators, since both cases can be treated almost simultaneously \cite{Fitzpatrick:2014oza,Lemos:2015awa}, but hereafter we focus only on the case $\NN=2$. 
The function $g_{\Delta,\ell}^{\Delta_{12},\Delta_{34}}(z,\bar{z})$ is the standard bosonic block for the decomposition of a correlation function with four distinct operators, defined in \eqref{eq:bosblock}. Although not immediately obvious, the bosonic block with shifted arguments in \eqref{eq:superblock} can be written as a finite sum of $g_{\Delta,\ell}^{0,0}(z,\bar{z})$ blocks, as expected from supersymmetry. The block reduces to $1$ for the identity exchange, \ie, $\Delta=\ell=0$.

The stress-tensor multiplet $\hat{\CC}_{0 (0,0)}$ corresponds to $\Delta=2$, $\ell=0$ in \eqref{eq:superblock}, and its OPE coefficient can be fixed using the Ward identities (see for example \cite{Beem:2014zpa}):
\be
\left|\lambda_{\phi \bar{\phi} \OO_{\Delta=2,\ell=0}}\right|^2 = \frac{\Delta_\phi^2 }{6 c}\,,
\label{eq:STOPEcoeff}
\ee
while long multiplets $\AA^{\Delta > \ell +2}_{0, 0, \ell}$ contribute as \eqref{eq:superblock} with $\Delta > \ell +2$.

When writing the crossing equations it will be useful to have the block expansion with a slightly different ordering
\be 
\langle \bar{\phi}_{-r}(x_1)  \phi_{r}(x_2) \phi_{r}(x_3) \bar{\phi}_{-r}(x_4)  \rangle = 
\frac{1}{x_{12}^{2\Delta_{\phi}} x_{34}^{2\Delta_{\phi}}} \sum_{\OO_{\Delta,\ell}} (-1)^{\ell} |\lambda_{ \phi  \bar{\phi} \OO}|^2\tilde{\GG}_{\Delta, \ell}(z,\bar{z})\,,
\label{eq:nonchiralbraid}
\ee
where the function $\tilde{\GG}_{\Delta, \ell}(z,\bar{z})(z,\bar{z})$ is defined as
\be 
\label{eq:superblockbraid}
\tilde{\GG}_{\Delta, \ell}(z,\bar{z}) =  (z \bar{z})^{-\frac{\NN}{2}} g_{\Delta + \NN,\ell}^{\NN,-\NN}(z,\bar{z})\, ,
\ee
and again we are only interested in the case $\NN=2$.

\subsubsection{Chiral channel}

The OPE selection rules of two identical $\NN=2$ chiral primary operators read \cite{Beem:2014zpa}
\be 
\phi_{r} \times \phi_{r} \sim   \EE_{2r}    + \CC_{0, 2r- 1 (j-1,j) } 
 + \BB_{1, 2r-1 (0,0)} + \CC_{\frac12, 2r- \frac32 (j-\frac12,j)} + \AA_{0, 2r-2 (j,j)}^{\Delta > 2+2r+2j}   \,,
\label{eq:selruleschiral}
\ee
where we already imposed Bose symmetry, and we assumed $\phi_r$ to be above the unitarity bound, \ie, $r >1$. If one considers different operators, or if $r=1$, additional multiplets are allowed to appear (see \eg, \cite{Lemos:2015awa}).
Chirality of $\phi_{r}$ requires each supermultiplet contributes with a single conformal family, and therefore the superblock decomposition contains only bosonic blocks:
\be
\label{eq:chiral}
\langle \phi_r(x_1) \phi_r(x_2) \bar{\phi}_{-r}(x_3) \bar{\phi}_{-r}(x_4) \rangle =  \sum_{\Delta,\ell} |\lambda_{\phi \phi \OO_{\Delta, \ell}}|^2 g^{0,0}_{\Delta,\ell}(z,\bar{z})\,.
\ee
Since we are considering the OPE between two identical $\phi_r$ multiplets, Bose symmetry requires 
the above sum to include only even $\ell$.
The precise contribution from each of the multiplets appearing in \eqref{eq:selruleschiral} is the following
\begin{equation}
\begin{alignedat}{4}
&{\AA_{0, 2r-2 (j,j)}}							&:\qquad &	g^{0,0}_{\Delta,\ell}\,,				\qquad &\Delta > 2+2r+\ell\,, \; \ell \; \mathrm{even}\,,\\
&{\CC_{\tfrac12, 2r-\tfrac32 (j-\tfrac12,j) }}	&:\qquad &	g^{0,0}_{\Delta=2r+\ell+2,\ell}\,,		\qquad & \ell \geqslant 2 \,,\;\ell\; \mathrm{even}\,, \\
&{\BB_{1, 2r-1 (0,0)}}							&:\qquad &	g^{0,0}_{\Delta=2r+2,\ell=0}\,,		\qquad &~\\
&{\CC_{0, 2r- 1 (j-1,j) }}						&:\qquad &	g^{0,0}_{\Delta=2r+\ell,\ell}\,,		\qquad & \ell \geqslant 2\,,\;\ell\; \mathrm{even}\,,\\
&{\EE_{2r}}										&:\qquad &	g^{0,0}_{\Delta=2r,\ell=0}\,,			\qquad &~ 
\end{alignedat}
\label{eq:chiralblocks}
\end{equation}
where $\ell= 2j$ is even (see \cite{Lemos:2015awa} for the contribution in the case of different operators).
While the short multiplets being exchanged in this channel have their dimensions fixed by supersymmetry, their OPE coefficients are not known. In fact, it is not even guaranteed all these multiplets are present as their physical meaning is not as clear as the short multiplets exchanged in the non-chiral channel. The $\EE_{2r}$ multiplet corresponds to an operator in the Coulomb branch chiral ring and therefore must be present in the $(A_1,A_2)$ theory, although the value of its OPE coefficient is not known. 
The scalar primary of the $\BB_{1, 2r-1 (0,0)}$ multiplet may be identified with a mixed branch chiral ring operator \cite{Argyres:2015ffa}, and since the $(A_1,A_2)$ has no mixed branch, one might expect this multiplet to be absent. We must point out, however, that the identification of this multiplet with the mixed branch is conjectural, it could be that the multiplet is present but it does not correspond to a flat direction \cite{Tachikawa:2013kta,Argyres:2015ffa}. 
Note that the contribution of the two short operators $\BB_{1, 2r-1 (0,0)}$ and $\CC_{\frac12, 2r- \frac32 (j-\frac12,j)}$ is identical to that of a long multiplet saturating the unitarity bound $\Delta= 2 + 2r + \ell$, as follows directly from the decomposition of the long multiplet when hitting the unitarity bound \cite{Dolan:2002zh}. On the other hand, the contribution of the short multiplets $\EE_{2r}$ and $\CC_{0 , 2r- 1 (j-1,j) } $ is isolated from the continuous spectrum of long operators by a gap; this will be relevant for the numerical analysis of section \ref{sec:numericsh0}.

\subsubsection{Crossing symmetry}

We are now ready to quote the crossing equations to be studied in the subsequent sections, where we recall only even spins are allowed in the $\phi \phi$ OPE,
\begin{subequations}
\be 
(z \zb)^{\Delta_\phi} \sum\limits_{\OO \in \phi \phi} |\lambda_{\phi \phi \OO}|^2 g^{0,0}_{\Delta_\OO, \ell_\OO} (1-z,1-\zb) = ((1-z)(1-\zb))^{\Delta_\phi }\sum\limits_{\OO \in \phi \bar\phi} |\lambda_{\phi \bar \phi \OO}|^2 (-1)^\ell \tilde{\GG}_{\Delta_\OO, \ell_\OO}(z,\zb) \,,
\label{eq:chiralnonchiral}
\ee   
\be 
((1-z)(1-\zb))^{\Delta_\phi} \sum\limits_{\OO \in \phi \bar\phi} |\lambda_{\phi \bar \phi \OO}|^2 \GG_{\Delta_\OO, \ell_\OO} (z,\zb) 
= (z \zb)^{\Delta_\phi} \sum\limits_{\OO \in \phi \bar\phi} |\lambda_{\phi \bar \phi \OO}|^2 \GG_{\Delta_\OO, \ell_\OO}(1-z,1-\zb)\,.
\label{eq:nonchiralnonchiral}
\ee 
\label{eq:crossingeqs}
\end{subequations}
The full system of equations comprises \eqref{eq:crossingeqs}  together with equation \eqref{eq:chiralnonchiral} with $z \to 1-z$ and $\zb \to 1-\zb$. These are collected in a form suitable for the numerical implementation in \eqref{eq:cross_numerics}.

\subsubsection{Numerical bootstrap}
\label{subsubsec:implementation}

In this short section we give details of the numerical implementation that will be necessary to understand the results of subsequent sections. Schematically, the final form of the crossing equations given in \eqref{eq:cross_numerics} is
\be 
|\lambda_{\OO^\star}|^2 \vec{V}_{\OO^\star} + \sum\limits_{\OO} |\lambda_{\OO}|^2 \vec{V}_{\OO} +\vec{V}_{\mathrm{fixed}} = 0\,.
\ee
Here $\OO^\star$ is a superconformal multiplet whose OPE coefficient we would like to bound numerically. 
The term $\vec{V}_{\mathrm{fixed}} $ encodes the contribution of the identity, or of the identity and stress tensor if we fix the central charge $c$, and is given in \eqref{eq:idandst}. 
OPE coefficient bounds are obtained using the SDPB solver of \cite{Simmons-Duffin:2015qma} to solve the following optimization problem
\be
\begin{split}
&\vec{\Psi}\cdot \vec{V}_{\OO} \geqslant 0 \,, \qquad \forall \; \OO\in \{\text{trial spectrum}\}\,,\\
&\vec{\Psi}\cdot \vec{V}_{\OO^\star} = \pm 1\,,  \\
&\mathrm{Maximize}\left(\vec{\Psi}\cdot \vec{V}_{\mathrm{fixed}}\right)\,,
\end{split}
\label{eq:optimiz}
\ee
where the minus sign in the second line can be consistently imposed at the same time as the first line, \textit{only} when the contribution of  $\OO^\star$ is isolated from the contribution of the remaining $\OO\in \{\text{trial spectrum}\}$ \cite{Poland:2011ey}.
As is standard in the bootstrap literature, we truncate the infinite-dimensional  functional as
\be  
\vec{\Psi} = \sum\limits_{m,n}^{m+n \leqslant\Lambda} \vec{\Psi}_{m,n} \partial_z^m \partial_{\zb}^n \Big\vert_{z= \zb=\frac12}\,.
\ee
The result of the extremization problem \eqref{eq:optimiz} provides a bound on the OPE coefficient of $\OO^\star$ as
\be 
\pm |\lambda_{\OO^\star}|^2 \leqslant -\mathrm{Max}\left(\Psi\cdot \vec{V}_{\mathrm{fixed}}\right)\,.
\ee
When the bound is saturated, there is a unique solution to the (truncated) crossing equations \cite{Poland:2010wg,ElShowk:2012hu}, with different extremization problems possibly leading to different solutions.\footnote{See, however, \cite{Behan:2017rca} for subtitles that arise when considering systems of mixed correlators.} 
At finite $\Lambda$, this corresponds to an approximate solution to the full crossing system, with the spectrum encoded in the extremal functional \cite{ElShowk:2012hu}.
We refer the reader to, \eg, \cite{Poland:2011ey,Simmons-Duffin:2015qma,Simmons-Duffin:2016gjk} for more technical details pertaining the numerical bootstrap.

\subsection{The Superconformal Index}

In the non-chiral channel the OPE coefficients of all short multiplets can be fixed, as they correspond either to the stress-tensor multiplet $\hat{\CC}_{0,0}$, which contributes as \eqref{eq:STOPEcoeff}, or to multiplets containing conserved currents of spin greater than two  $\hat{\CC}_{0,\ell >0}$ which are absent in interacting theories \cite{Maldacena:2011jn,Alba:2013yda}.  
In the chiral OPE selection rules \eqref{eq:selruleschiral}, there appear short multiplets whose dimensions are fixed, but whose OPE coefficients are not known, and could even be absent in the $(A_1,A_2)$ theory. 
To access the spectrum of short operators, a useful quantity is the superconformal index \cite{Kinney:2005ej,Gadde:2011uv,Rastelli:2014jja}. The index is the most general invariant that counts, with signs, short multiplets up to multiplets that can recombine to form a long multiplet. Because short multiplets that have the right quantum numbers to recombine in a long multiplet give zero contribution, the index has some intrinsic ambiguities. Relevant to us are the following recombination rules
\be 
\begin{split}
\AA^{2R+r+2 j_2 +2}_{R,r(j_1,j_2)} &\to \CC_{R,r(j_1,j_2)} \oplus \CC_{R+\frac{1}{2},r+\frac{1}{2}(j_1-\frac{1}{2},j_2)}\,,\\
\AA^{2R+r+2 j_2 +2}_{R,r(0,j_2)} &\to \CC_{R,r(0,j_2)} \oplus \BB_{R+1,r+\frac{1}{2}(0,j_2)}\,,
\end{split}
\label{eq:recombination}
\ee
where the latter can be seen as a special case of the former with the identification $\CC_{R,r(-\frac{1}{2},j_2)} = \BB_{R+\frac{1}{2},r(0,j_2)}$.
This means that, while the multiplets $\BB_{1, 2r-1 (0,0)}$, $\CC_{0, 2r- 1 (j-1,j) }$, and $\CC_{\frac12, 2r- \frac32 (j-\frac12,j)}$ appearing in \eqref{eq:selruleschiral} contribute to the superconformal index, we can only see if these multiplets appear modulo pairs of the type \eqref{eq:recombination}. Note that the $\EE_r$, $\bar{\EE}_{-r}$ multiplets themselves can never recombine.
 
The \emph{full} superconformal index of the $(A_1,A_2)$  theory has been computed in \cite{Maruyoshi:2016tqk} using a $4d$ $\NN=1$ Lagrangian theory which flows to the $(A_1,A_2)$ theory in the IR.\footnote{Various limits of the superconformal index of Argyres-Douglas theories had been  obtained before in \cite{Buican:2015ina,Buican:2015tda,Cordova:2015nma,Song:2015wta}.}
The final expression for the index is given in integral form in equation (16) of \cite{Maruyoshi:2016tqk}, which we use to gather information about the spectrum of short multiplets in the theory. Expanding said expression we find, unsurprisingly, that the Coulomb branch chiral ring operators $\EE_{k \frac65}$, with integer $k\geqslant1$ are present (since we explored the index in an expansion this was only checked for low values of $k$). In section \ref{sec:OPEbounds} we shall bound, from above and from below, the OPE coefficient of the $\EE_{\frac{12}5}$ operator appearing in the $\EE_{\frac65}$ self OPE (see \eqref{eq:Ebound}).
From the index we also find that the operator $\CC_{0,\frac75(0,1)}$  is present in the spectrum, which corresponds to the leading-twist contribution of spin two in \eqref{eq:chiralblocks}. We shall recover this result in section \ref{sec:OPEbounds}, since we find a non-zero lower bound for its OPE coefficient (see \eqref{eq:Clbound}).
Finally, we find no contributions arising from a  $\BB_{1, \frac75(0,0)}$ multiplet implying that, modulo 
the recombination ambiguities of \eqref{eq:recombination}, it is absent in the $(A_1,A_2)$ theory. This is consistent with the expectation that such multiplets may be identified with the existence of a mixed branch \cite{Argyres:2015ffa}. Although this is not conclusive evidence for the absence of this multiplet, we take it as an indication that if there are multiple solutions to the crossing equations for $r_0=\frac65$ and $c=\tfrac{11}{30}$, then the one corresponding to the $(A_1,A_2)$ theory likely has the $\BB_{1, \frac75(0,0)}$ multiplet absent.
\section{Numerical results}
\label{sec:numericsh0}

In this section we attempt to zoom in on the $(A_1,A_2)$ Argyres-Douglas theory reviewed in the previous section, following up on the numerical analysis of the Coulomb branch presented in \cite{Beem:2014zpa,Lemos:2015awa}, where the landscape of theories with one or two Coulomb branch chiral ring operators was explored. 

An interesting question that was left open in \cite{Lemos:2015awa} was whether the $(A_1,A_2)$ theory saturates the numerical lower bounds on the central charge $c$.
While it has been established analytically that this theory has the lowest possible central charge among interacting $\NN=2$ SCFTs \cite{Liendo:2015ofa}, 
this does not preclude solutions to the crossing equations \eqref{eq:crossingeqs} for $r_0=\frac65$ and values of $c$ smaller than $\frac{11}{30}$. 
The central charge bounds from \cite{Beem:2014zpa,Lemos:2015awa} were obtained for $\Lambda \leqslant 22$, and there was no particularly clear trend that would allow for an extrapolation to $\Lambda \to \infty$.

In this work we present improved numerical results, with extrapolations consistent with, but not definite proof of, the saturation of the $c$-bound by the $(A_1,A_2)$ SCFT.
Moreover, our results seem to imply that, even if there is more than one crossing symmetric four-point function for $r_0=\frac65$ and $c=\tfrac{11}{30}$, these solutions do not differ by much as far as some observables are concerned, and can be used as an approximation to the low-lying spectrum of the $(A_1,A_2)$ theory.
In particular, we are able to obtain the first
predictions for unprotected OPE coefficients in the form of \emph{true} upper and lower bounds for OPE coefficients, together with conservative extrapolations for $\Lambda \to \infty$. In addition, we also estimate the value of the lowest-twist unprotected long multiplets appearing in the non-chiral OPE. In this section we focus on the lowest spin operators, but numerical results for larger spins are presented in section \ref{sec:anlytical}, where we compare them to  estimates arising from the Lorentzian inversion formula of \cite{Caron-Huot:2017vep} adapted to the supersymmetric case.

\subsection{Central charge bound}
\label{sec:cbound}

\begin{figure}[htbp!]
             \begin{center}           
              \includegraphics[scale=0.35]{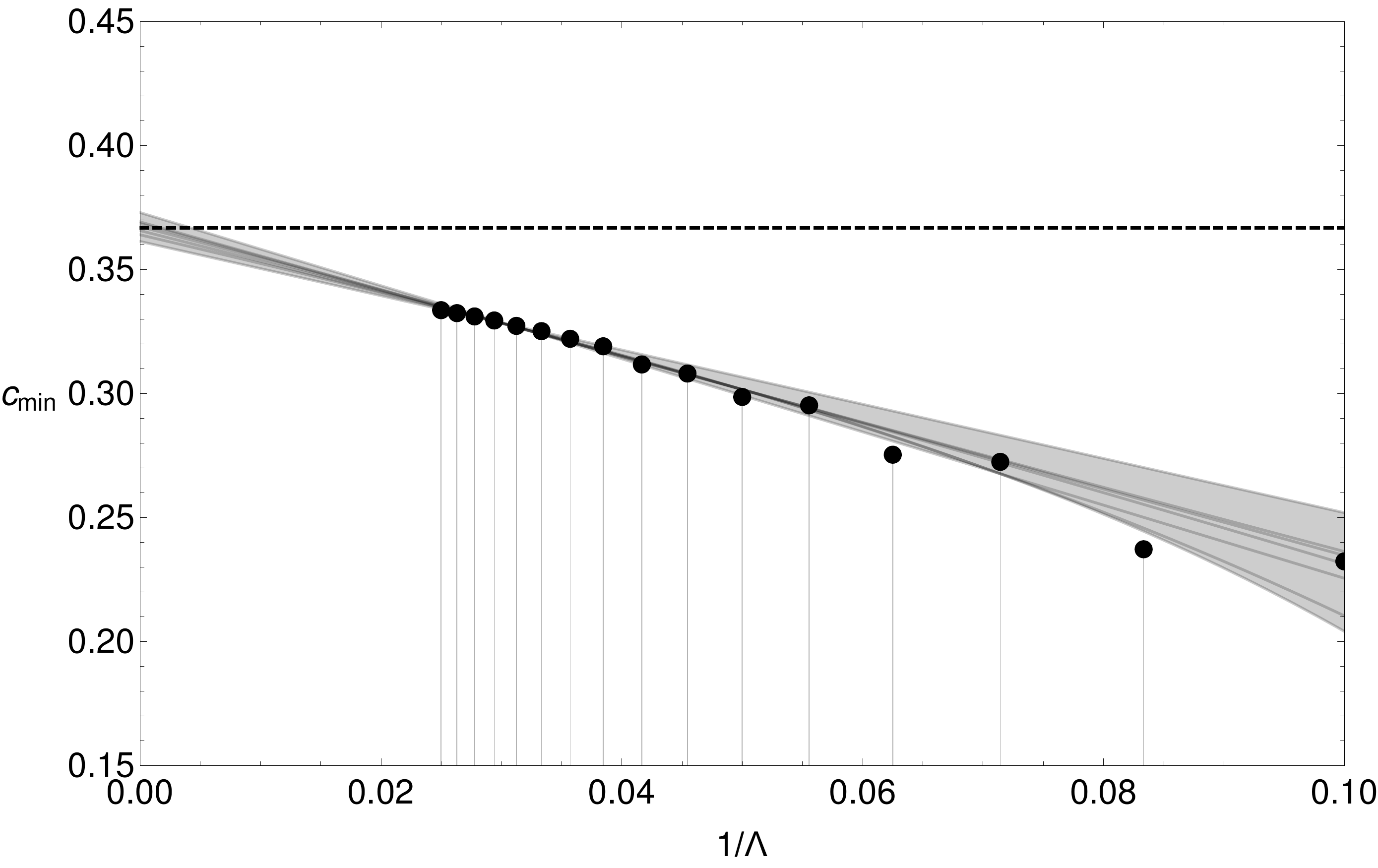}
              \caption{Numerical lower bound (black dots) on the central charge of theories with an $\NN=2$ chiral operator of dimension $r_{0}=\frac65$ as a function of the inverse cutoff $\Lambda$. The lines correspond to various extrapolations to infinitely many derivatives, and the horizontal dashed line marks $c=\tfrac{11}{30}$ -- the central charge of the $(A_1,A_2)$ SCFT.}
              \label{Fig:c_min}
            \end{center}
\end{figure}
Our first task is to obtain numerical lower bounds on the central charge $c_{min}(\Lambda)$, for fixed $r_0=\frac65$, as a function of the cutoff $\Lambda$.
The resulting bound $c_{min}(\Lambda)$ is shown in figure \ref{Fig:c_min}, together with various different extrapolations to $\Lambda \to \infty$. 
While the results are consistent with the bound converging to $c_{min}=\tfrac{11}{30}$ (the dashed line in figure \ref{Fig:c_min}), they are still not conclusive enough. In what follows we will be agnostic about the $\Lambda \to \infty$ fate of the $c$-bound, and concentrate on a region around $c\sim\tfrac{11}{30}$ in an attempt to estimate the CFT data of the $(A_1,A_2)$ theory.

\subsection{OPE coefficient bounds}
\label{sec:OPEbounds}

We now concentrate on OPE-coefficient bounds for the different short multiplets appearing in the chiral channel, for varying $c \geqslant c_{min}(\Lambda)$, and external dimension fixed to $r_0=\frac65$. In particular, we obtain an upper bound for the OPE coefficient of the $\BB_{1,\frac75(0,0)}$ multiplet, 
and both \emph{lower and upper} bounds (see discussion in subsection \ref{subsubsec:implementation}) for the coefficients of the $\EE_{\frac{12}5}$ and $\mathcal{C}_{0,\frac75\left(\frac{\ell }{2}-1,\frac{\ell }{2}\right)}$ multiplets.

For fixed $\Lambda$, there is a unique solution to the truncated crossing equations \eqref{eq:crossingeqs} at $c=c_{min}(\Lambda)$ \cite{Poland:2010wg,ElShowk:2012hu}, and indeed we will see below that upper and lower bounds (when available) coincide. As already discussed, it is plausible that $c_{min}(\Lambda) \to \tfrac{11}{30}$ as $\Lambda \to \infty$, and so in this limit the meeting point of upper and lower bounds would be at $c=\frac{11}{30} \simeq 0.367$.

An important subtlety in all the plots that follow is that we cannot fix the central charge exactly: each time we quote a value of $c$, the corresponding plot captures values \textit{less or equal} than the given number. This follows from the fact that we allow for a continuum of long multiplets with dimensions consistent with the unitarity bounds; for the non-chiral channel this means long multiplets with $\Delta > \ell +2$ \eqref{eq:selrulesnonchiral}. However, as is clear from the superconformal blocks \eqref{eq:superblock}, the contribution of a long multiplet at the unitarity bound mimics the contribution of a conserved current $\hat{\CC}_{0,\ell}$. This has two important consequences. First, we cannot restrict ourselves to interacting theories, because it is not possible to set to zero the OPE coefficient of the conserved currents of spin greater than two ($\hat{\CC}_{0,\ell \geqslant 1}$), without imposing a gap on the spectrum of all long multiplets. Second, even if we fix the central charge according to \eqref{eq:STOPEcoeff}, a long multiplet at the unitarity bound with an arbitrary (positive) coefficient, will increase the value of the $\lambda_{\phi \bar{\phi} \OO_{\Delta=2, \ell=0}}$ coefficient, which means that we are really allowing for all central charges smaller than the fixed value. This implies that a given bound can only get weaker as $c$ is increased, and explains the flatness of some of the bounds presented below.

\subsubsection*{OPE coefficient bound for $\BB_{1,\frac75(0,0)}$}

\begin{figure}[htb!]
             \begin{center}           
              \includegraphics[scale=0.363]{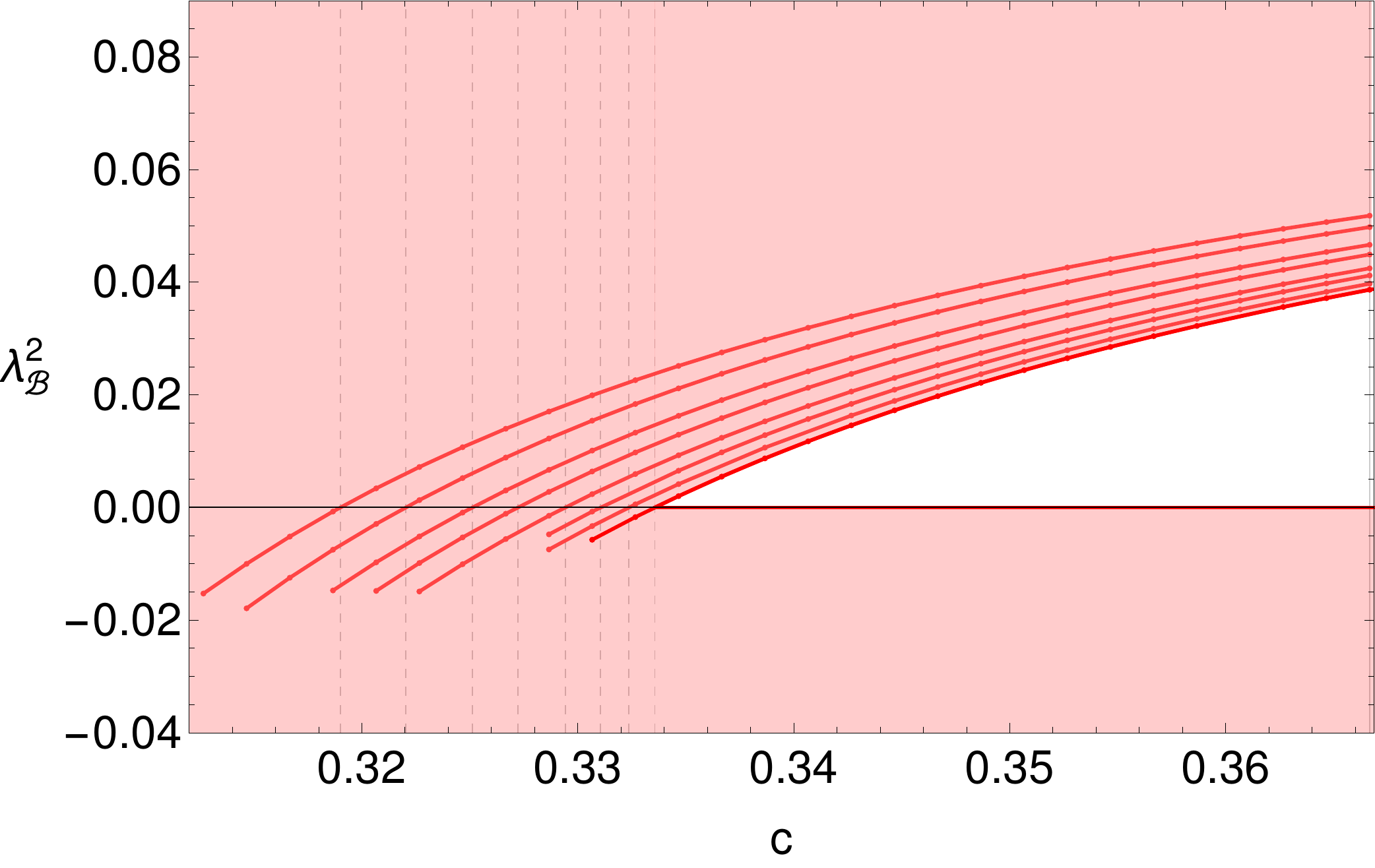}
               \includegraphics[scale=0.362]{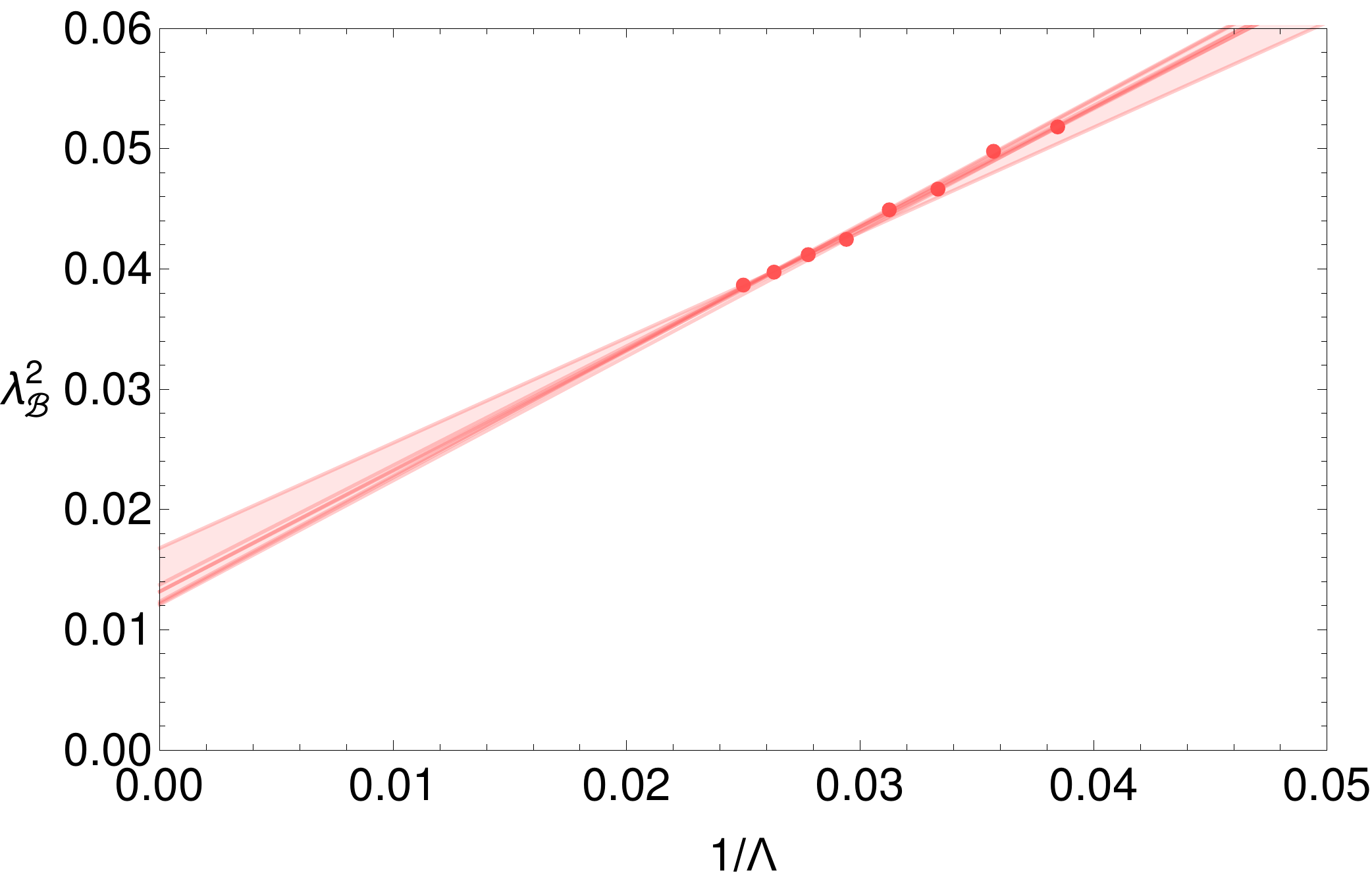}
              \caption{Numerical upper bound on the OPE coefficient squared of the operator $\BB_{1,\frac75(0,0)}$ appearing in the chiral channel for $\Lambda=26,\ldots,40$, and external dimension $r_0=\frac65$. Left: Upper bound on the OPE coefficient for different values of the central charge, with the strongest bound corresponding to $\Lambda=40$; the dashed lines mark the minimum central charge as extracted from figure \ref{Fig:c_min} for each cutoff $\Lambda$, and the solid line marks $c=\tfrac{11}{30}$. Right: Bound on the OPE coefficient for $c=\tfrac{11}{30}$ as a function of the inverse cutoff $\Lambda$, together with various extrapolations to infinitely many derivatives.}
              \label{Fig:Bbound}
            \end{center}
\end{figure}

Let us first consider the OPE coefficient squared of $\BB_{1,\frac75(0,0)}$. A numerical upper bound, as a function of the central charge, and for fixed external dimension $r_0=\frac65$, is shown on the left-hand side of figure~\ref{Fig:Bbound} for various values of the cutoff $\Lambda$. For each value of $\Lambda$, the upper bound vanishes for $c=c_{min}(\Lambda)$ (marked by the dashed vertical lines in the figure), and becomes negative for $c<c_{min}(\Lambda)$, implying there is no unitary solution to the crossing equations. This is consistent with what was found for $\Lambda=12$ in \cite{Lemos:2015awa}, and suggests this operator is responsible for the existence of the central charge bound.
Since such a multiplet is associated with the mixed branch, and the $(A_1,A_2)$ theory has no mixed branch, it 
would be natural to expect its absence to be a feature of the four-point function of the $(A_1,A_2)$ theory.
However, as can be seen on the right-hand side of figure \ref{Fig:Bbound},
the numerical results appear to leave room for solutions to crossing with a small value of this OPE coefficient, as it is not clear if the upper bound will converge to zero as $\Lambda \to \infty$.
If there is more than one solution, it is plausible that the one corresponding to the $(A_1,A_2)$ theory is one in which $\BB_{1,\frac75(0,0)}$ has zero OPE coefficient. We should point out though, that the absence of a mixed branch does not guarantee that the aforementioned multiplet is absent, as it is possible that such a multiplet is present, but one cannot give it a vev and thus no mixed branch exists \cite{Tachikawa:2013kta,Argyres:2015ffa}.

\subsubsection*{OPE coefficient bound for \texorpdfstring{$\EE_{\frac{12}5}$}{E2r0 OPE coefficient bound}}

\begin{figure}[htb!]
             \begin{center}           
              \includegraphics[scale=0.363]{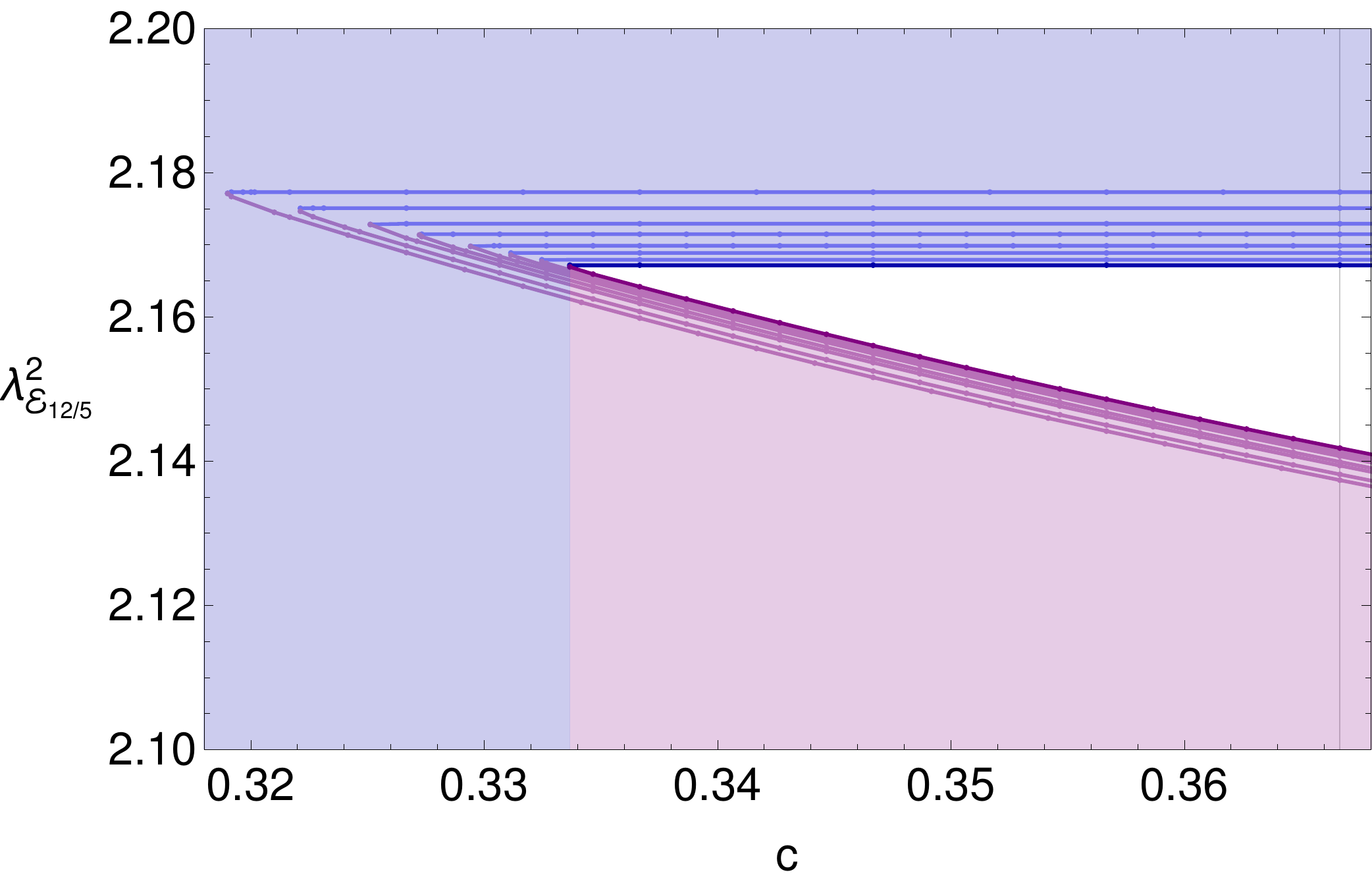}
              \includegraphics[scale=0.362]{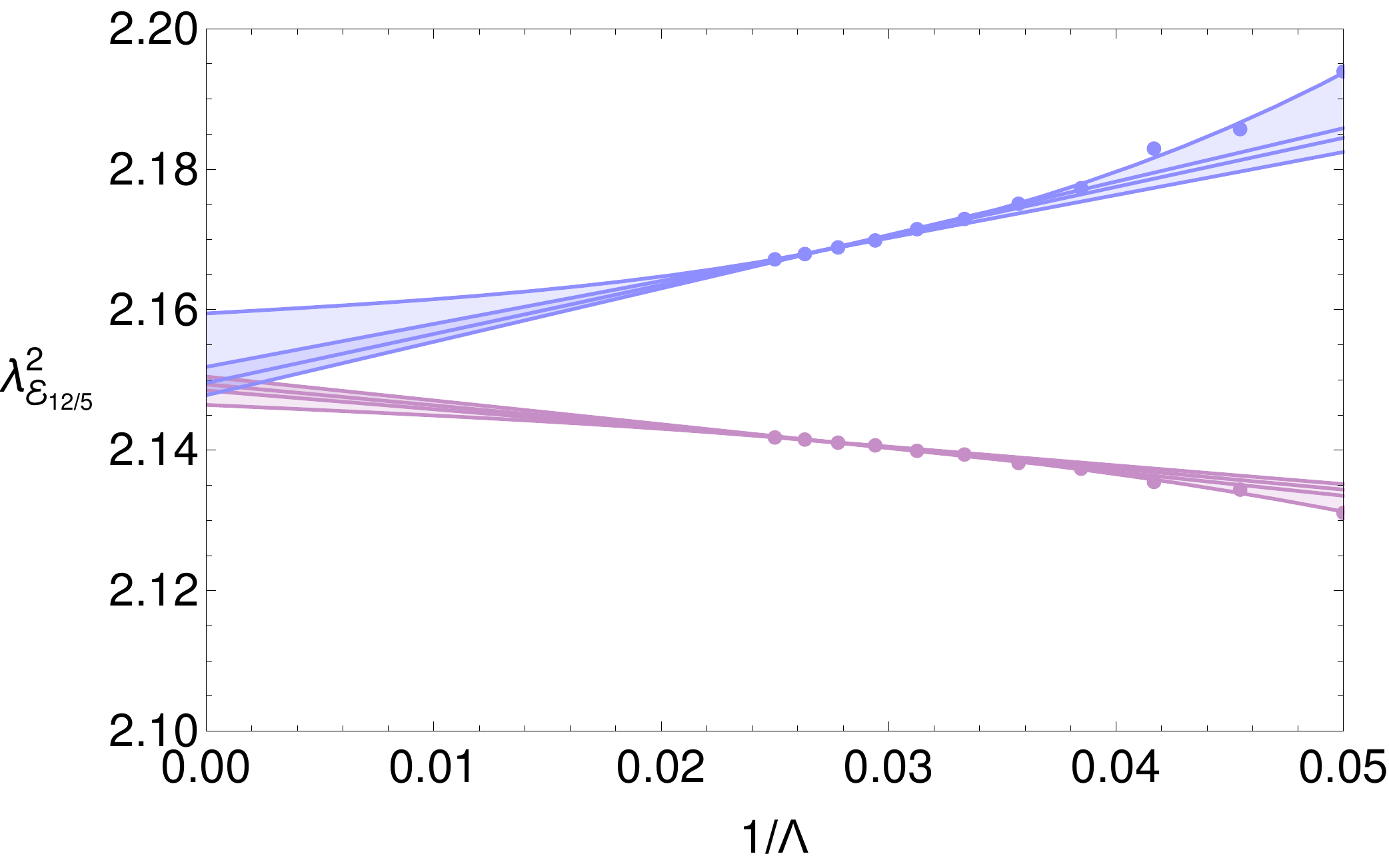}
              \caption{Numerical upper and lower bounds on the OPE coefficient squared of the chiral operator $\EE_{\frac{12}5}$ for increasing number of derivatives and external dimension $r_0=\frac65$. Left: Bounds on the OPE coefficient for different values of the central charge, with cutoffs $\Lambda=26,\ldots,40$, the vertical line marks $c=\tfrac{11}{30}$. Right: Various extrapolations of the lower and upper bounds at $c=\tfrac{11}{30}$ for infinite $\Lambda$.}
              \label{Fig:Ebound}
            \end{center}
\end{figure}

Turning to the OPE coefficient of the Coulomb branch chiral ring operator $\EE_{\frac{12}5}$, we can now place upper and lower bounds  as a function of the central charge. We present the results on the left-hand side of figure~\ref{Fig:Ebound} for several values of $\Lambda$.

As already discussed, the plots in this section allow for all central charges $c\geqslant c_{\mathrm{fixed}}$, since a gap in the spectrum of spin zero long multiplets is not imposed. This explains the flatness of the upper bound: solutions to crossing saturating it can effectively have central charges equal to $c_{min}(\Lambda)$.\footnote{A natural solution would be to impose small gaps in the spectrum of long multiplets, this removes the conserved currents of spin greater than two and fixes the central charge. However, we have no intuition on the size of these gaps, not even for spin zero, as there is no understanding of the number of non-supersymmetry preserving relevant deformations. We 
experimented imposing that the spectrum of long multiplets obeys $\Delta \geqslant 2+\epsilon + \ell$ for various small values of $\epsilon$, and although the upper bound gets stronger than that of figure \ref{Fig:Ebound}, it varies smoothly with $\epsilon$ and thus there is no justification to pick any specific value. The lower bound, on the other hand, shows a much smaller dependence on $\epsilon$.} 
The lower bound, however, must be saturated by theories with central charge equal to the fixed value.
At $c_{min}(\Lambda)$ the upper and lower bounds coincide, fixing a unique value of the OPE coefficient, and as $c$ is increased a wider range of values, and distinct solutions to crossing, are allowed. We show the allowed range, for $c=\tfrac{11}{30}$, as a function of $1/\Lambda$ on the right-hand side of figure~\ref{Fig:Ebound}. The lines correspond to different extrapolations through (subsets) of the data points, and the shaded region aims to give an idea of where the bounds are converging to. If $c_{min}(\Lambda \to \infty) = \tfrac{11}{30}$, then the upper and lower bound should converge to the same value, which is not ruled out by the extrapolations. In any case, our results indicate that the OPE coefficient of $\EE_{\frac{12}5}$ is constrained to a narrow range.

We have thus obtained the following 
\emph{rigorous} bounds for the value of this OPE coefficient in the $(A_1,A_2)$ theory:
\be
2.1418 \leqslant \lambda_{\EE_{\frac{12}5}}^2 \leqslant 2.1672 \,, \qquad \mathrm{for}\; \Lambda=40\,.
\label{eq:Ebound}
\ee
Furthermore, the most conservative of the extrapolations presented in figure \ref{Fig:Ebound} gives
\be
2.146 \lesssim \lambda_{\EE_{\frac{12}5}}^2 \lesssim 2.159 \,, \qquad \text{extrapolated for}\; \Lambda\to \infty\,.
\label{eq:Eboundextrapol}
\ee

\subsubsection*{OPE coefficient bounds for $\CC_{0,\frac75(0,1)}$  and $\CC_{0,\frac75(1,2)}$}

\begin{figure}[htbp!]
\begin{center}      
	\begin{subfigure}[t]{0.5\textwidth}
	\centering        
	\includegraphics[scale=0.362]{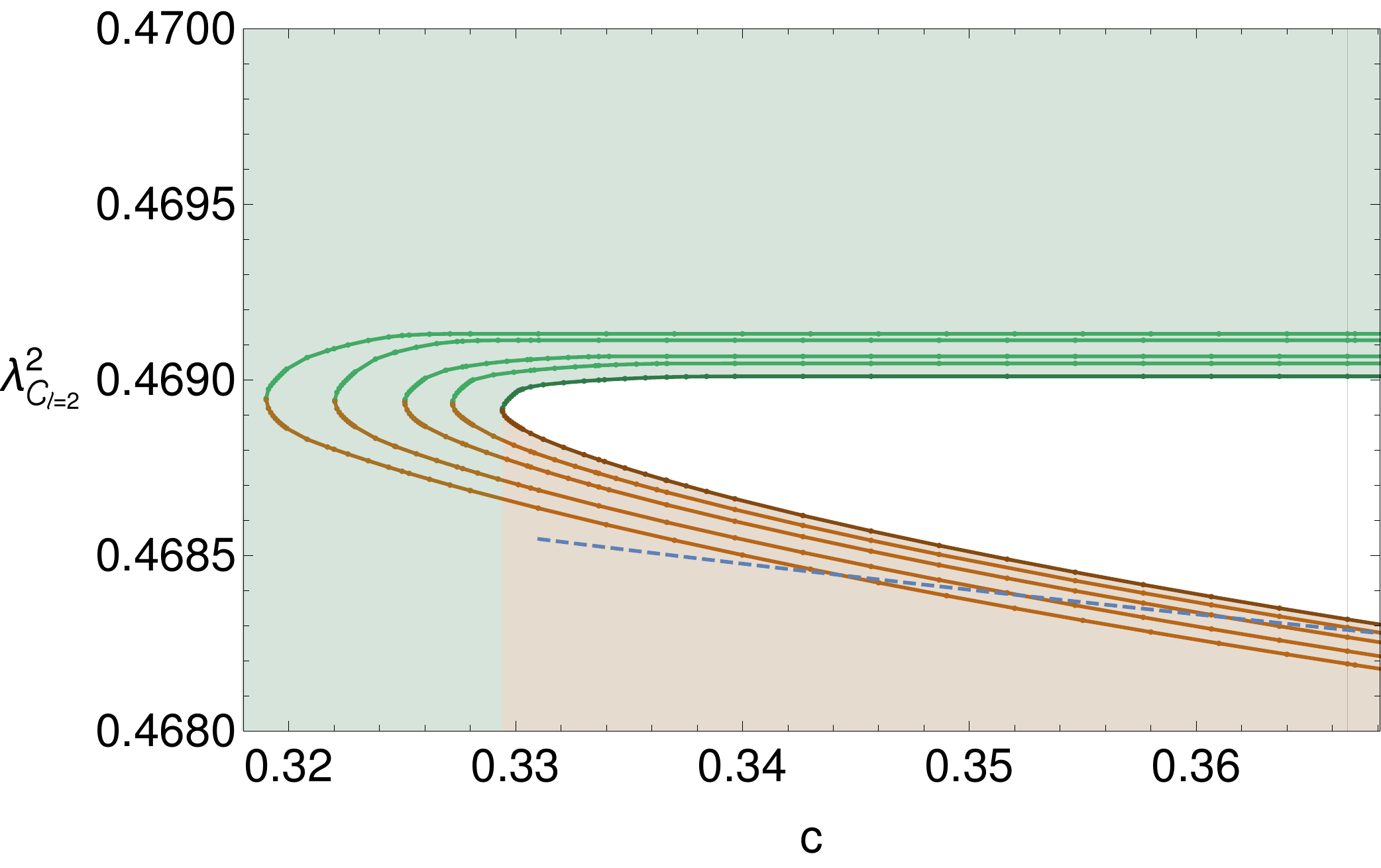}
	\caption{$\CC_{0,\frac75(0,1)}$}
    \label{Fig:Cl2}
    \end{subfigure}~
    \begin{subfigure}[t]{0.5\textwidth}
    \centering        
    \includegraphics[scale=0.362]{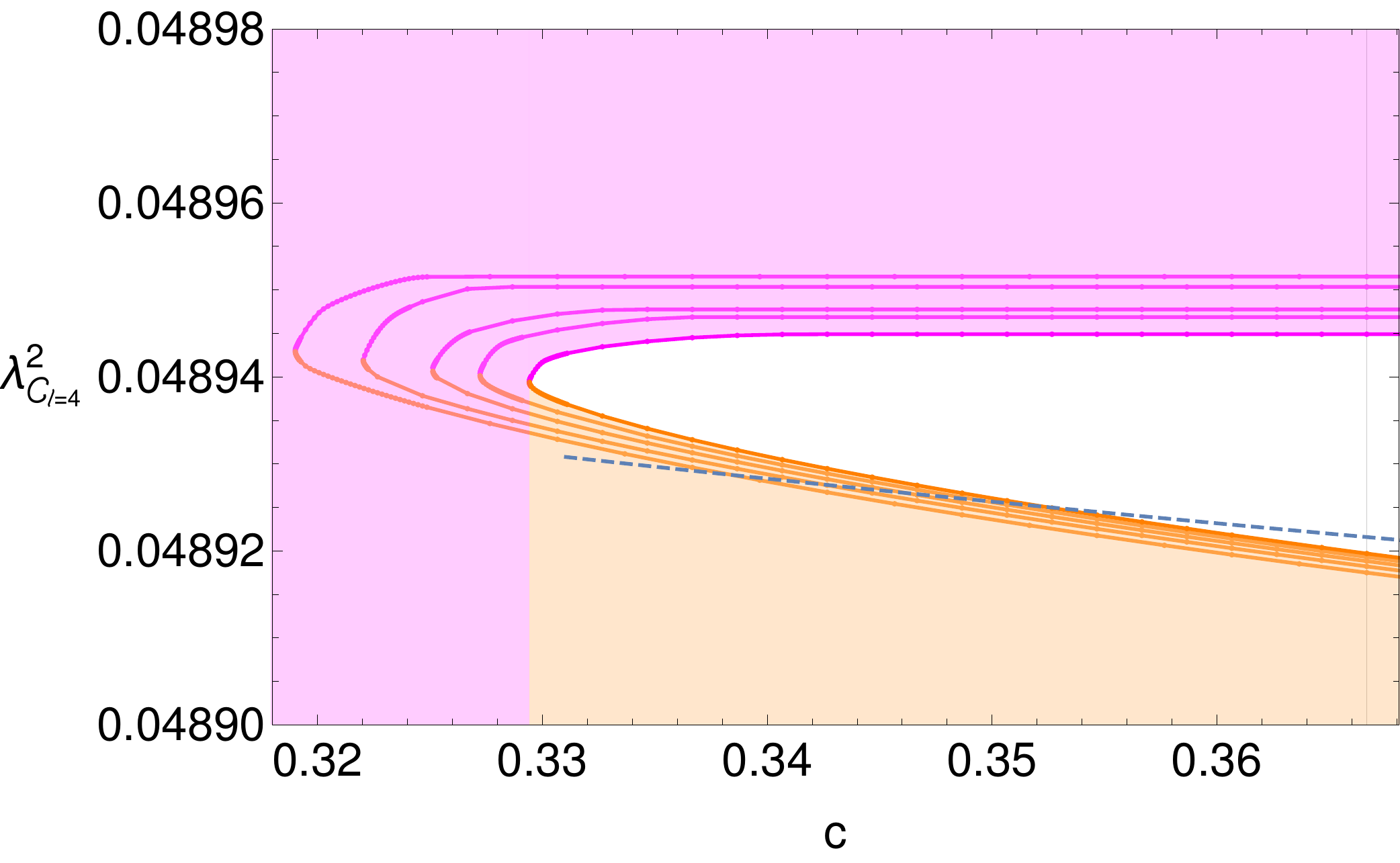}
    \caption{$\CC_{0,\frac75(1,2)}$}
    \label{Fig:Cl4}
    \end{subfigure}
\end{center}
\caption{Numerical upper and lower bounds on the OPE coefficient squared of the chiral channel multiplet $\mathcal{C}_{0,\frac75\left(\frac{\ell }{2}-1,\frac{\ell }{2}\right)}$, for $\ell=2,4$, and external dimension $r_0=\frac65$. The bounds were obtained for cutoffs $\Lambda=26,\ldots,34$ and the vertical line marks $c=\tfrac{11}{30}$. The dashed line corresponds to the value obtained from the Lorentzian inversion formula of \cite{Caron-Huot:2017vep} applied to the chiral channel and using as input only the exchange of the identity and stress-tensor superblocks in the non-chiral channel, and thus valid for sufficiently large $\ell$ (see section \ref{sec:chiral_inv} for more details).}
\label{Fig:Cl24}
\end{figure}

Let us now focus on the $\mathcal{C}_{0,\frac75\left(\frac{\ell }{2}-1,\frac{\ell }{2}\right)}$ family of multiplets. Like in the $\EE_{\frac{12}5}$ case, upper and lower bounds are possible thanks to the gap that separates these $\CC$-type multiplets from the continuum of long operators. The bounds for $\ell=2,4$,  as a function of $c$, are shown in figures \ref{Fig:Cl2} and \ref{Fig:Cl4} respectively, while bounds for higher values of $\ell$ can be found in figure \ref{Fig:ClCH}, for fixed $c=\tfrac{11}{30}$.
The dashed lines in figure \ref{Fig:Cl24} are estimates of the OPE coefficient valid for sufficiently large $\ell$, that will be discussed in detail in section \ref{sec:chiral_inv}.
Similarly to the $\EE_{\frac{12}5}$ multiplet, the OPE coefficients of these multiplets in  the $(A_1,A_2)$ theory are constrained to lie in a narrow range:
\be
0.46831 \leqslant \lambda_{\CC_{0,\frac75, (0,1)}}^2 \leqslant 0.46901 \,, \qquad 
0.048919 \leqslant \lambda_{\CC_{0,\frac75, (1,2)}}^2 \leqslant 0.048945 \,,  \qquad
 \text{for}\; \Lambda=34\,.
 \label{eq:Clbound}
\ee

The upper bounds in figure \ref{Fig:Cl24} now show a mild dependence on the central charge, and so we can compare the extrapolations of the upper and lower bounds at $c=\tfrac{11}{30}$ with the extrapolation of the value of the OPE coefficient for the unique solution at $c_{min}(\Lambda)$. Like  before, the extrapolations (not shown) do not rule out that $c_{min} \to \tfrac{11}{30}$ as $\Lambda \to \infty$. As visible in figure \ref{Fig:Cl24}, the value of the OPE coefficient at $c_{min}(\Lambda)$ (the meeting point) shows a very mild dependence on $\Lambda$,  unlike the OPE coefficient of $\EE_{\frac{12}5}$, we can therefore obtain the following estimates 
\be 
\begin{split}
0.4687 & \lesssim \lambda_{\CC_{0,\frac75(0,1)}}^2 (c= c_{min}(\Lambda)) \lesssim 0.4688 \,, \\
0.04892 & \lesssim \lambda_{\CC_{0,\frac75(1,2)}}^2 (c= c_{min}(\Lambda))   \lesssim 0.04894\,, \\
\end{split}\qquad
 \text{extrapolated for }\; \Lambda\to \infty\,.
 \label{eq:Clboundextrapol}
\ee

\subsection{Dimensions of unprotected operators}
\label{sec:dimensions}

\begin{figure}[htbp!]
\begin{center}
	\begin{subfigure}[t]{0.5\textwidth}        
	\includegraphics[scale=0.365]{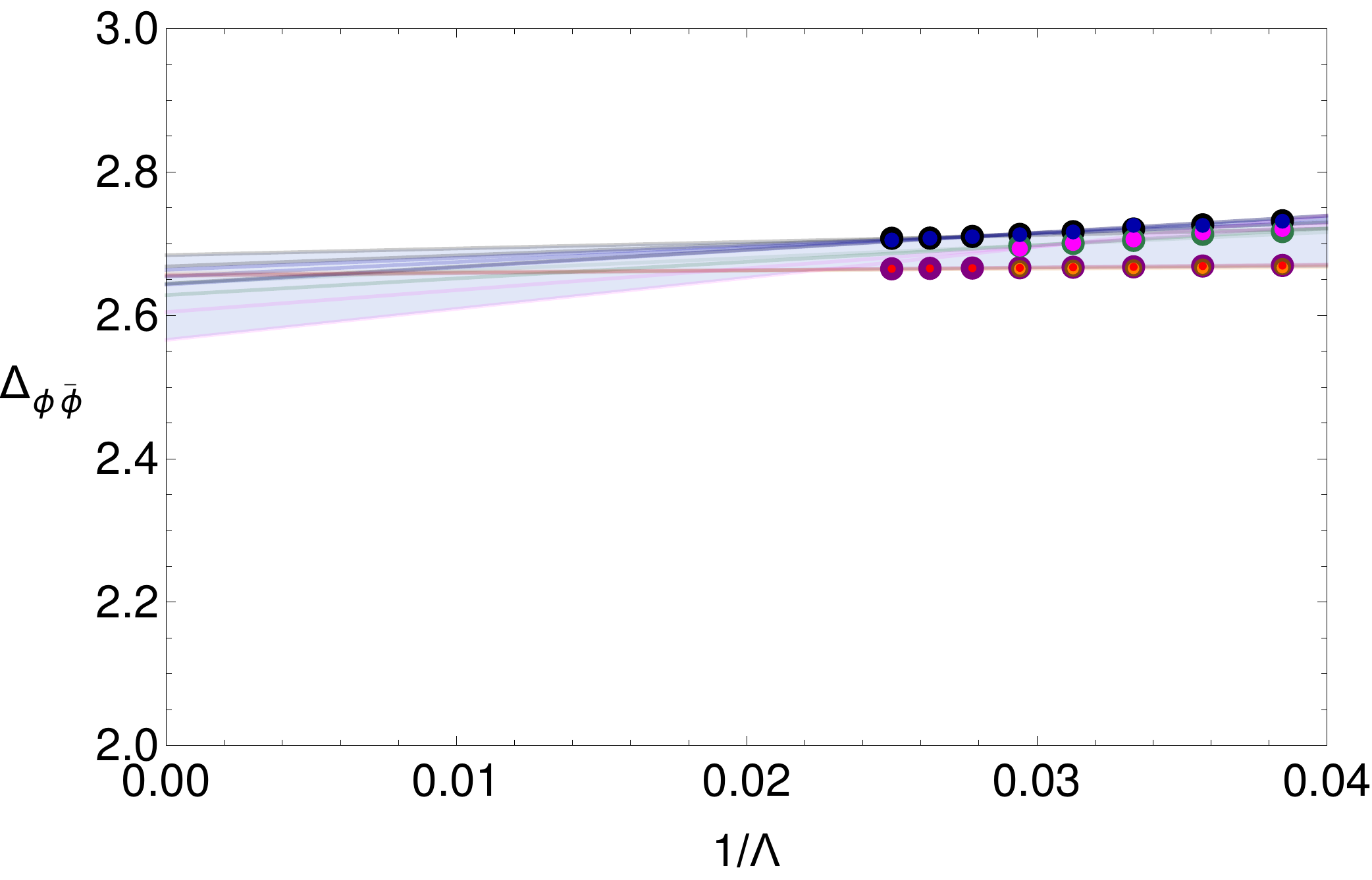}
	\caption{Dimension of first long multiplet in the non-chiral channel arising from different bounds.}
	\label{Fig:firstnonchirallong}
	\end{subfigure}~
	\begin{subfigure}[t]{0.5\textwidth}
	\includegraphics[scale=0.365]{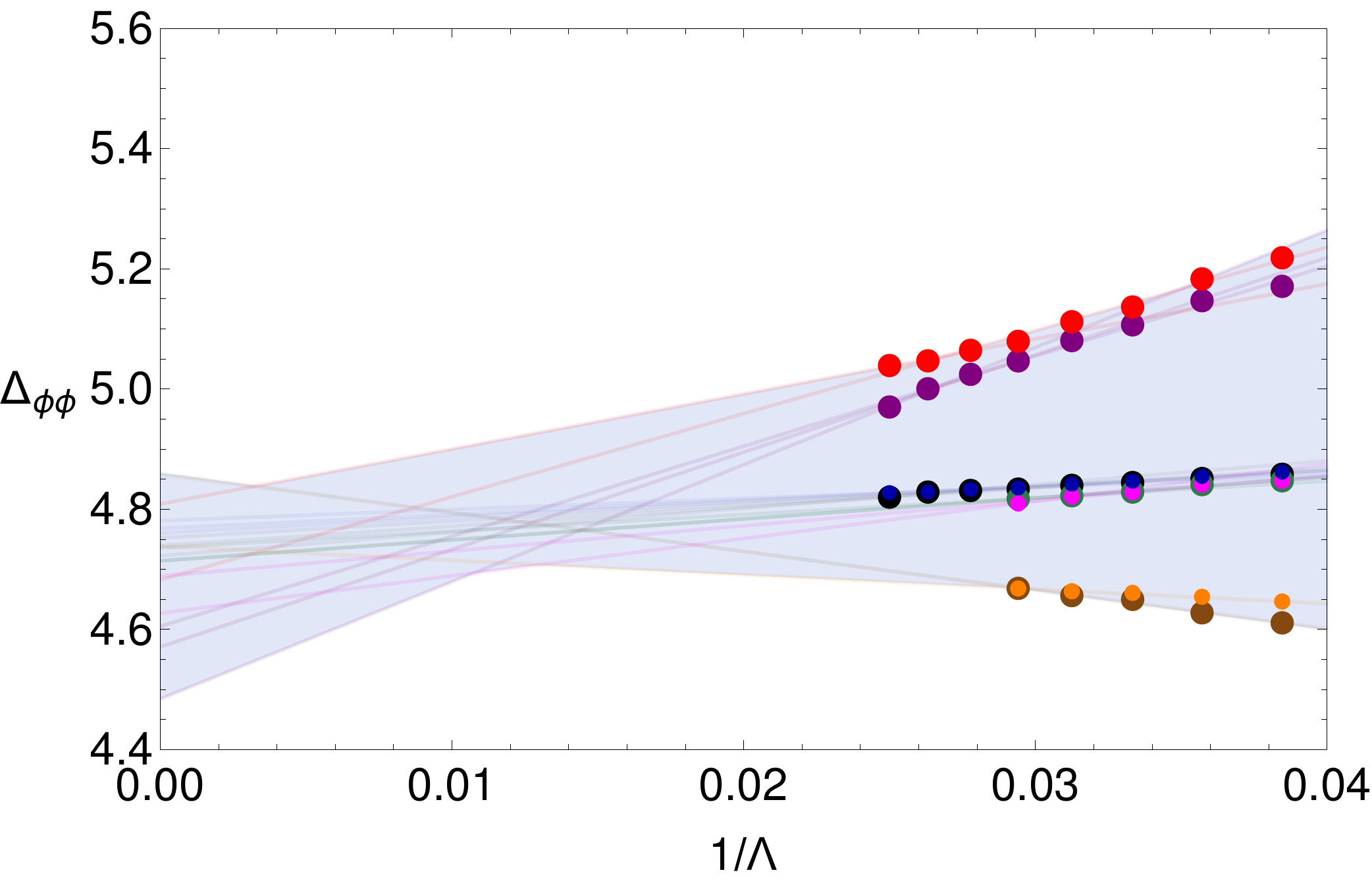}
	\caption{Dimension of first long multiplet in the chiral channel arising from different bounds.}
	\label{Fig:firstchirallong}
	\end{subfigure}
\end{center}
\caption{Numerical estimates for the first scalar long operator in the non-chiral (a) and chiral (b) channels obtained from the functionals of figures \ref{Fig:c_min}, \ref{Fig:Bbound}, \ref{Fig:Ebound}, and \ref{Fig:Cl24}. The data points are color-coded according to the bound the extremal functional was extracted from, and in the cases where the bounds are plotted as a function of $c$, the functional for $c=\tfrac{11}{30}$ was used. The lines give an estimate of the extrapolation to infinitely many derivatives.}
\end{figure}

Finally, we estimate dimensions of unprotected long operators. In \cite{Beem:2014zpa,Lemos:2015awa} numerical upper bounds on the dimensions of the first long in the non-chiral and chiral channels were obtained, for various values of $c$ and  $r_0$. The best bound obtained in \cite{Beem:2014zpa} for the dimension of the first scalar long operator in the  non-chiral channel reads $\Delta_{\phi \bar{\phi}} \leqslant 2.68$, for $\Lambda=18$, $r_0=\frac65$ and $c\leqslant\tfrac{11}{30}$. On the other hand, the bound obtained for the first scalar long operator in the chiral channel was very weak and converged too slowly without further assumptions (figure~2 of \cite{Lemos:2015awa}). Removing the $\BB_{1,\frac75(0,0)}$ multiplet this bound improved to $\Delta_{\phi \phi} \leqslant 4.93 $ for $\Lambda=20$, $r_0=\frac65$, and did not appear to depend on $c$ \cite{Lemos:2015awa}.

Here, instead, we extract the dimensions of the first long $\AA_{0,0,\ell}^{\Delta_{\phi\bar{\phi}}}$ and $\AA_{0,\frac25,\ell}^{\Delta_{\phi\phi}}$ multiplets in the approximate solutions to crossing saturating the various bounds presented above. 
The results for $\ell=0$, and various values of the cutoff $\Lambda$, are given in figure \ref{Fig:firstnonchirallong} for  $\Delta_{\phi \bar{\phi}}$, and in figure \ref{Fig:firstchirallong} for $\Delta_{\phi \phi}$.
The dimensions were extracted from the extremal functionals \cite{ElShowk:2012hu} of figures \ref{Fig:Bbound}--\ref{Fig:Cl24} for $c=\tfrac{11}{30}$, and from figure \ref{Fig:c_min}, and are color coded according to the bound they came from. Note that these are the dimensions of the first long operator present in each of the extremal solutions, and are not rigorous upper bounds, which would require a different extremization problem. The lines in the figures show various extrapolations of the dimensions to $\Lambda \to \infty$.

The estimates of the dimensions of the lowest-lying long multiplet in the non-chiral channel appear all consistent with each other, even at finite $\Lambda$. This implies that, even if the various extremization problems are solved by different solutions to the crossing equations, these solutions do not differ by much as far as $\Delta_{\phi \bar{\phi}}$ is concerned, and we can take the spread of the values as an estimate for the uncertainty in the value of this dimension. Then, conservative extrapolations for $\Lambda \to \infty$ of the values coming from the various functionals give
\be 
2.56 \lesssim \Delta_{\phi \bar{\phi}} \lesssim 2.68 \,, \qquad \text{from the extrapolations as } \Lambda \to \infty\,,
\label{eq:dimextrapol}
\ee
for $\Delta_{\phi \bar{\phi}}$ in the $(A_1,A_2)$ theory.
Similarly, the dimensions of the leading twist non-chiral operators with spin $\ell >0$, obtained from the various extremal functionals of figures \ref{Fig:c_min}--\ref{Fig:Cl24}, are shown in figure \ref{Fig:anomCH} for $\Lambda=34$. We will comment on these results in section \ref{sec:nonchiral_inv}.

Less coherent are the results for $\Delta_{\phi \phi}$, the values  extracted from the extremal functionals of figures \ref{Fig:c_min}--\ref{Fig:Cl24} look very different for finite $\Lambda$, and the extrapolations are not conclusive. This is shown in  figure \ref{Fig:firstchirallong}.
Since the dimensions obtained are so disparate, it is not clear we can get any meaningful estimate for this operator in the $(A_1,A_2)$ theory.

\section{Inverting the OPEs}
\label{sec:anlytical}

In recent years, starting from \cite{Fitzpatrick:2012yx,Komargodski:2012ek}, there has been much progress in understanding the large spin spectrum of CFTs by studying analytically the crossing equations in a Lorentzian limit.
In $d >2$, by studying 
the four-point function $\langle \phi_1(x_1) \phi_1(x_2) \phi_2(x_3) \phi_2(x_4) \rangle$ in the lightcone limit, the authors of \cite{Fitzpatrick:2012yx,Komargodski:2012ek} found that in the $t-$channel there must be exchanged, in a distributional sense, double-twist operators, \ie, operators whose dimensions approach $\Delta_1 + \Delta_2 + \ell$ for $\ell \to \infty$, and whose OPE coefficients tend to the values of generalized free field theory. Corrections to these dimensions and OPE coefficients, or rather weighted averages of these quantities, in a large spin expansion can be obtained in terms of the leading twist operators exchanged in the $s-$channel.\footnote{Some of the steps taken in the derivations \cite{Fitzpatrick:2012yx,Komargodski:2012ek} rely on intuitive assumptions, some of which have started to be put on a firm footing in \cite{Qiao:2017xif}.} 
Assuming the existence of individual operators close to the average values, this procedure was set up to systematically compute the OPE coefficients and dimensions of the double-twist operators in an asymptotic expansion in the inverse spin \cite{Alday:2015ewa}.
The lightcone limit of the crossing equations has been used to constrain the large spin spectrum of various CFTs with different global symmetries and supersymmetries \cite{Fitzpatrick:2012yx,Komargodski:2012ek,Alday:2013cwa,Fitzpatrick:2014vua,Vos:2014pqa,Fitzpatrick:2015qma,Kaviraj:2015cxa,Alday:2015eya,Kaviraj:2015xsa,Alday:2015ota,Beem:2015aoa,1612.03891,1603.05150,1611.01500,1612.00696,1510.08091,1602.04928,1511.08025}.
Remarkably, this has resulted in predictions for OPE coefficients and anomalous dimensions of operators that match the numerical results down to spin two \cite{Simmons-Duffin:2016wlq, Alday:2015ota}.

Recently, the work of \cite{Caron-Huot:2017vep} has explained this agreement, by showing that the spectrum organizes in analytic families.\footnote{We thank Marco Meineri for many discussions on \cite{Caron-Huot:2017vep}.} There, a ``Lorentzian'' inversion formula for the $s-$channel OPE of a given correlator was obtained, with the crucial feature that the result of the inversion is a function that is analytic in spin (a function valid only for spin greater than one). 
This established, for sufficiently large spin, the existence of each individual double-twist operator.
The inversion formula explained the organization of the spectrum, and allows one to compute individual OPE coefficients and anomalous dimensions, avoiding the asymptotic expansions, and obtaining the coefficients instead of averages.

\medskip

Motivated by the success of \cite{Simmons-Duffin:2016wlq}, we take the first steps towards a systematic analysis of the $(A_1,A_2)$ theory for large spin, considering both crossing equations \eqref{eq:crossingeqs}.
We apply the inversion formula obtained in \cite{Caron-Huot:2017vep}  to invert the chiral \eqref{eq:chiral} and non-chiral \eqref{eq:nonchiral} OPEs. The block decomposition of the former happens to be simply a decomposition in bosonic blocks \eqref{eq:chiralblocks}, and thus the inversion formula directly applies. The latter has a decomposition in superblocks, but as we shall see, the formula can still be applied, although we must work as if we had a correlator of unequal external operators. The only required modifications will be on the spin down to which the formula holds, and on what the crossed-channel decompositions are.
Since the numerical results in supersymmetric theories are not yet at the level of accuracy of the $3d$ Ising model, we refrain from using numerical data as input to the analysis, and instead compare the large-spin estimates coming from the inversion formula with the numerical results. The only input we give is the exchange of the identity and stress-tensor supermultiplet in the non-chiral OPE, and thus find results that are good estimates for sufficiently large spin. We find a reasonable agreement between the OPE coefficients of the leading-twist short operators exchanged in the chiral channel, and the analytical estimate, already for low spin, see figure \ref{Fig:ClCH}. Our analysis also shows that the anomalous dimensions of the double-twist operators in the non-chiral channel, arising from the stress tensor exchange, are small, and we confirm this by matching to numerical estimates of these dimensions obtained from the bounds of section \ref{sec:numericsh0} (see figure \ref{Fig:anomCH}).

We start with a brief summary of the results of \cite{Caron-Huot:2017vep} relevant for our purposes and refer the reader to that reference for further details. 
Starting from  the four-point function of unequal scalar operators,
\be 
\langle \OO_1(x_1)\OO_2(x_2)\OO_3(x_3)\OO_4(x_4) \rangle =\frac{1}{|x_{12}|^{\Delta_1+\Delta_2}|x_{34}|^{\Delta_3+\Delta_4}} \left|\frac{x_{14}}{x_{24}}\right|^{\Delta_{21}}  \left|\frac{x_{14}}{x_{13}}\right|^{\Delta_{34}} \GG(z, \zb)\,,
\label{eq:CHfourpoint}
\ee
the main result is a ``Lorentzian'' inversion formula for the $s-$channel decomposition of $\GG(z,\zb)$ in conformal blocks,
\be 
\GG(z, \zb) = \sum\limits_{\Delta,\ell} \lambda_{\Delta,\ell}^2 g^{\Delta_{12},\Delta_{34}}_{\Delta,\ell}(z, \zb)\,.
\label{eq:Gdecbos}
\ee
The OPE coefficients $\lambda_{\Delta,\ell}^2$ (for $\ell >1$) of the above decomposition \eqref{eq:Gdecbos} are then encoded in the residues of a function $c(\ell,\Delta)$  that is analytic in spin, contrasting with the one that can be obtained from a Euclidean inversion of the OPE. The condition $\ell >1$ arises during the contour manipulations needed to go from the  Euclidean inversion of the OPE, valid only for integer $\ell$, to the ``Lorentzian'' formula of \cite{Caron-Huot:2017vep}. This condition requires looking at the $t-$ and $u-$ channels to bound the growth of $\GG(z,\zb)$  in a particular region, and it is valid for any unitary CFT.
The function $c(\ell, \Delta) $ receives contributions from the $t-$ and  $u-$channels, with the even and odd spin operators defining two independent trajectories, as 
\be 
c(\ell, \Delta) = c^t (\ell, \Delta) + (-1)^\ell c^u(\ell, \Delta)\,,
\label{eq:ctandcu}
\ee
where $c^t$ and $c^u$ are defined in (3.20) of \cite{Caron-Huot:2017vep}.

The poles of $c(\ell, \Delta)$ in $\Delta$, at fixed $\ell$, encode the dimensions of the operators in the theory, with the residues giving the  OPE coefficients.\footnote{In some cases the residues need to be corrected as discussed in (3.9) of \cite{Caron-Huot:2017vep}, but for the computations carried out in this section this correction is not needed.} As described in section 3.2 of \cite{Caron-Huot:2017vep}, if one is interested only in getting the poles and residues, the inversion formula can be written as
\be
\begin{split}
c^t(\ell,\Delta)\big\vert_{\mathrm{poles}} &= \int\limits_0^1 \frac{\d z}{2z} z^{\frac{\ell-\Delta}{2}} \left( \sum\limits_{m=0}^{\infty} z^m \sum\limits_{k=-m}^{m} B_{\ell,\Delta}^{(m,k)} C^t(z, \ell+\Delta+2k)\right)\,, \\
C^t(z, \beta)&= \int\limits_z^1 \d \zb \frac{(1-\zb)^{\frac{\Delta_{21}+\Delta_{34}}{2}}}{\zb^2} \kappa_{\beta} k_\beta^{\Delta_{12}, \Delta_{34}}(\zb) \dDisc\left[\GG(z,\zb)\right]\,,
\end{split}
\label{eq:CHgeneratingbos}
\ee
and similarly for $c^u(\ell,\Delta)$.
Here  $\dDisc$ denotes the double-discontinuity of the function, 
$k_\beta^{\Delta_{12}, \Delta_{34}}(\zb)$ is defined in equation \eqref{eq:bosblock}, and 
\be 
\kappa_{\beta}= \frac{\Gamma \left(\frac{\beta -\Delta_{21}}{2}\right) \Gamma \left(\frac{\Delta_{21}+\beta }{2}\right) \Gamma \left(\frac{\beta-\Delta_{34} }{2}\right) \Gamma \left(\frac{\Delta_{34} + \beta }{2}\right)}{2 \pi ^2 \Gamma (\beta -1) \Gamma (\beta )} \,.
\ee
The $z \to 0$ limit of the block in \eqref{eq:CHgeneratingbos} gave the collinear block $k_\beta^{\Delta_{12}, \Delta_{34}}(\zb)$ which does not take into account all descendants, these are instead taken into account by the functions $B_{\ell,\Delta}^{(m,k)}$, as discussed in \cite{Caron-Huot:2017vep}. Since we shall focus only on leading twist operators we do not need to subtract descendants and thus do not need these functions, apart from $B_{\ell,\Delta}^{(0,0)}=1$.
A term $z^{\frac{\tau(\Delta+\ell)}{2}}$ in the bracketed term in \eqref{eq:CHgeneratingbos}  implies there exists a pole at $\Delta-\ell = \tau(\Delta+\ell)$, with its residue, taken at fixed $\ell$, providing the OPE coefficient;  see \cite{Caron-Huot:2017vep} for more details.

\subsection{Inverting the chiral OPE}
\label{sec:chiral_inv}

The Lorentzian inversion formula obtained in \cite{Caron-Huot:2017vep} can be directly applied to invert the $s-$channel OPE of the correlator \eqref{eq:chiral},
\be 
\langle \phi(x_1) \phi(x_2) \bar{\phi}(x_3) \bar{\phi}(x_4) \rangle = \frac{1}{x_{12}^{2 \Delta_\phi}x_{34}^{2 \Delta_\phi}} \sum\limits_{\OO_{\Delta,\ell}} |\lambda_{\phi \phi \OO_{\Delta,\ell}}|^2 g_{\Delta,\ell}^{0,0}(u,v)\,,
\label{eq:chiralcorr}
\ee 
as it is exactly of the form \eqref{eq:CHfourpoint}, with $\GG(z,\zb)$ admitting a decomposition in bosonic blocks, with $\Delta_{12}=\Delta_{34}=0$. We can thus apply \eqref{eq:ctandcu} directly, with the $t-$ and $u-$channel decompositions as dictated by crossing symmetry of \eqref{eq:chiralcorr}. Since the operators at point one and two are identical, the $u-$ and $t-$ channels give identical contributions, and thus only even spins appear in the $s-$channel OPE of \eqref{eq:chiralcorr}, precisely in agreement with Bose symmetry.

We now want to make use of the generating functional \eqref{eq:CHgeneratingbos}, to obtain the dimensions and OPE coefficients of the $s-$channel operators (at least for large enough spin) by providing information about the $t-$channel decomposition.
For large spin (that is large $\beta$) the leading contributions in \eqref{eq:CHgeneratingbos} come from the $\zb \to 1$ limit of the integrand, with the leading contribution corresponding to the lowest twist operators exchanged in the $t-$channel.
The $t-$channel decompositions are given by the non-chiral OPE, as follows from \eqref{eq:chiralnonchiral}.
From \eqref{eq:selrulesnonchiral} we see that, after the identity, the leading contribution comes from the superconformal multiplet of the stress tensor $\hat{\CC}_{0,0}$, and, since we are interested on interacting theories, there is no other contribution with the same twist. The next contributions will arise from long multiplets, for which we only currently have the numerical estimates for their dimensions  obtained in section \ref{sec:dimensions}. At large spin the contributions of one of these operators of twist $\tau$ behaves as $\ell^{-\tau}$ \cite{Fitzpatrick:2012yx,  Caron-Huot:2017vep}.\footnote{Here we are using the bosonic results of \cite{Fitzpatrick:2012yx,  Caron-Huot:2017vep}, while we have a superblock contribution at twist $\tau$. However, decomposing the superblock in bosonic blocks we find a finite number of bosonic blocks with twist $\tau$ together with a finite number of higher twist, and so the presence  of the superblock will only modify the coefficient of the leading behavior for large $\ell$, which is unimportant for our point here.}
The leading twist operators have been estimated numerically from the various extremal functionals obtained in section \ref{sec:numericsh0}. The leading spin zero operator could have twist as low as $\tau \sim 2.5$, while the higher spin operators (figure \ref{Fig:anomCH}) all appear to have twists close to $2r_0=2.4$, with small corrections depending on the spin. This is to be compared with the contribution of the stress tensor with exactly $\tau=2$.
For sufficiently large spin the contributions of long multiplets are subleading,
so in what follows we shall consider only the stress-tensor and identity exchanges in the $t-$ and $u-$channels.

The identity and stress tensor contribute to \eqref{eq:CHgeneratingbos} according to the crossing equation \eqref{eq:chiralnonchiral}, with the identity contributing as $|\lambda_{\phi \bar{\phi} 1}|^2 \tilde{\GG}_{0,0}(u,v)=1$. The stress-tensor superblock is given by \eqref{eq:superblockbraid} with $\Delta=2$, $\ell=0$, and with OPE coefficient given by \eqref{eq:STOPEcoeff}, and we  find
\be 
C^t(z,\beta) \supset \int\limits_0^1 \frac{\d \zb}{\zb^2} \kappa_{\beta} k_\beta^{\Delta_{12}, \Delta_{34}}(\zb) \dDisc\left[\frac{(z \zb)^{r_0}}{((1-z) (1-\zb))^{r_0}} \left( 1+ |\lambda_{\phi \bar{\phi} \hat{\CC}_{0,0}}|^2 \tilde{\GG}_{2,0}(1-z,1-\zb)\right)\right]\,.
\label{eq:nonchiral_lead}
\ee

From the identity contribution, which is the leading one for large spin, we recover the existence of double-twist operators $[\phi\phi]_{m,\ell}$ (see for example section 4.2 of \cite{Caron-Huot:2017vep}), namely operators with dimensions approaching
\be 
\Delta_{[\phi\phi]_{m,\ell}} \underset{\ell \gg 1}{\longrightarrow} 2 r_0+2m + \ell\,, \qquad \ell \; \mathrm{even}\,,
\ee
and with OPE coefficients approaching those of generalized free field theory,
\be 
\lambda_{\mathrm{gfft}}^2 = \frac{\left((-1)^{\ell }+1\right) \left(\left(r_0-1\right)_m\right){}^2 \left(\left(r_0\right)_{m+\ell }\right){}^2}{m! \ell ! \left(m+2 r_0-3\right)_m (\ell +2)_m \left(m+\ell +2
   r_0-2\right)_m \left(2 m+\ell +2 r_0-1\right)_{\ell }}\,,
 \label{eq:gfftOPEcoeff}
\ee
at large spin. In \eqref{eq:gfftOPEcoeff} $(a)_b$ denotes the Pochhammer symbol.

To compute the leading correction to these dimensions and OPE coefficients at large spin we take into account the contribution of the stress-tensor multiplet to the OPE. 
To do so we take the $z \to 0 $ limit of \eqref{eq:nonchiral_lead}; as pointed in \cite{Caron-Huot:2017vep}, the correct procedure should be to subtract a known sum, such that the limit $z \to 0 $ commutes with the infinite sum over $t-$channel primaries.
However, when anomalous dimensions are small this procedure gives small corrections to the naive one of taking a series expansion in $z$ and extracting anomalous dimensions from the terms proportional to $\log z$  (the generating function should have $z^{\gamma/2} \approx 1+ \tfrac{1}{2} \gamma \, \log z + \ldots $) and corrections to OPE coefficients from the terms without $\log z$. For the case considered below the situation is even better as the anomalous dimensions of the operators we are interested in vanish.
Taking the small $z$ limit, the first observation is that anomalous dimensions, \ie, log-terms, only come with a power of $z^{\Delta_\phi+2}$, and thus only the operators $[\phi\phi]_{m \geqslant 2,\ell}$ acquire an anomalous dimension. This is consistent with the fact that from the block decomposition \eqref{eq:chiralblocks} we identify the double-twist operators with $m=0,1$ as short multiplets, $\CC_{0, 2r_0- 1 (\frac{\ell}2-1,\frac{\ell}2) } $ and $\CC_{\frac12, 2r_0- \frac32 (\frac{\ell}2-\frac12,\frac{\ell}2)}$ respectively, whose dimensions are protected.\footnote{We assume $\ell\geqslant 2$ here since the inversion formula is not guaranteed to converge for $\ell=0$.}

\begin{figure}[htb!]
             \begin{center}           
              \includegraphics[scale=0.35]{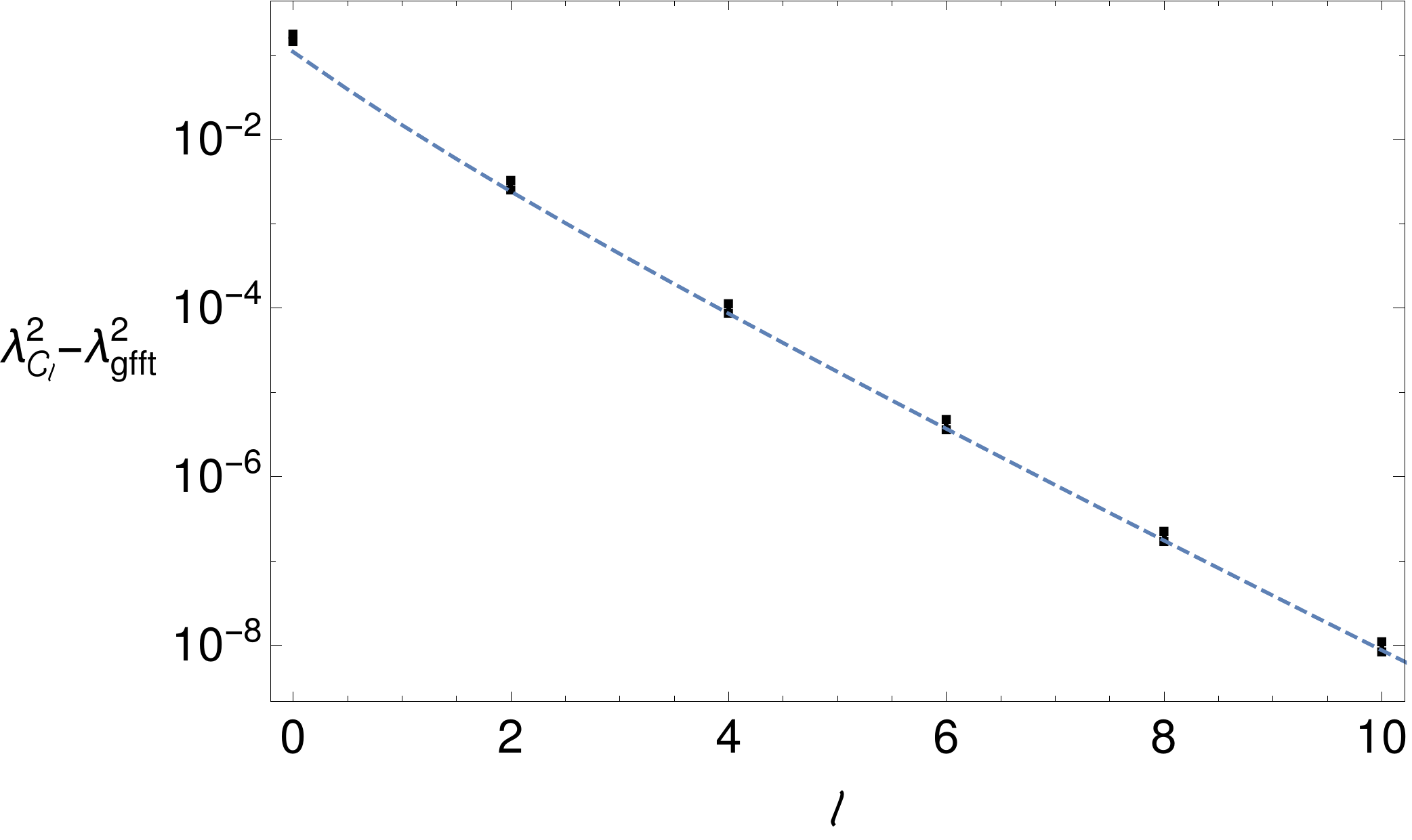} \; \includegraphics[scale=0.35]{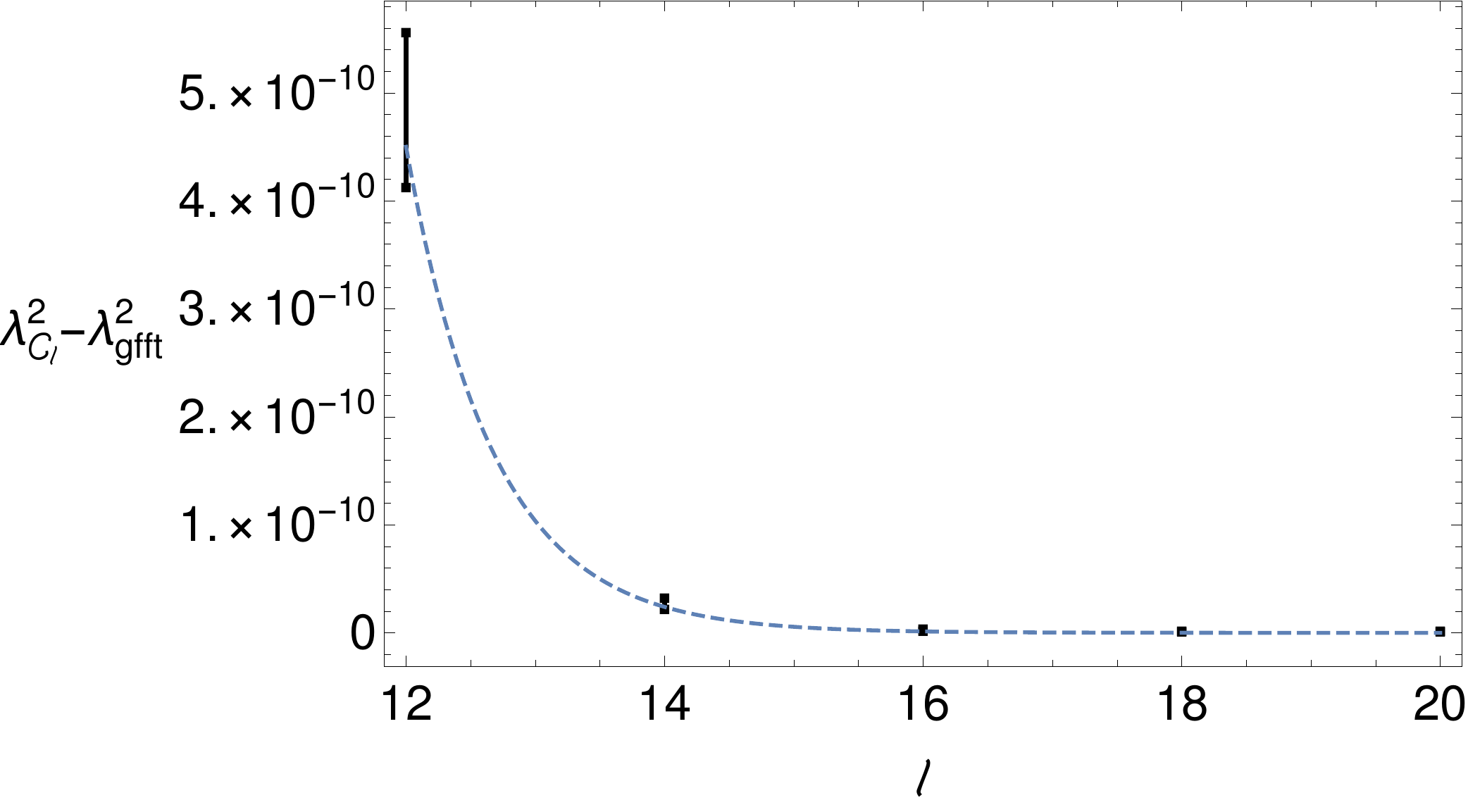}
              \caption{Comparison between the numerical bounds on the OPE coefficient squared of the leading twist operators in the chiral channel and the results from  the inversion formula \eqref{eq:nonchiral_lead}, for $c=\tfrac{11}{30}$ and external dimension $r_0=\frac65$. The black boxes mark the numerically allowed range for the squared OPE coefficients of the $\mathcal{C}_{0,\frac75,\left(\frac{\ell }{2}-1,\frac{\ell }{2}\right)}$ operators for different values of $\ell$ (the $\ell=0$ operator should be interpreted as $\EE_{\frac{12}5}$) and with $\Lambda=36$. The dashed line shows the result of equation \eqref{eq:nonchiral_lead}, where we considered only the contribution of the identity and stress-tensor operators in the non-chiral channel, and thus is an approximate result for sufficiently large spin. The formula \eqref{eq:nonchiral_lead} is not guaranteed to be valid for $\ell=0$, and the results here are just shown as an illustration.}
              \label{Fig:ClCH}
            \end{center}
\end{figure}

We can now compute corrections to the OPE coefficient of the $\mathcal{C}_{0,2 r_0-1,\left(\frac{\ell }{2}-1,\frac{\ell }{2}\right)}$ operators, for $r_0=\frac65$ and $c=\tfrac{11}{30}$, from \eqref{eq:nonchiral_lead}.
The result is plotted in figure \ref{Fig:ClCH}, where we performed the integral in \eqref{eq:nonchiral_lead} numerically (after taking the leading $z$ term), together with the numerical upper and lower bonds on the OPE coefficients, obtained in section \ref{sec:OPEbounds}.\footnote{Note that by the usual lightcone methods \cite{Fitzpatrick:2012yx,Komargodski:2012ek,Alday:2015ewa}, we could obtain an asymptotic expansion in $\tfrac{1}{\ell}$ of the correction to the generalized free field theory OPE coefficients \eqref{eq:gfftOPEcoeff} arising from the stress tensor exchange.  By considering the contributions of the stress tensor superblock to \eqref{eq:CHgeneratingbos} we are effectively re-summing the lightcone expansion to all orders.}
The results for $\ell=0$ (where the multiplet becomes an $\EE_{2r_0}$) are also shown even though the formula is only guaranteed to be valid for $\ell \geqslant 2$.\footnote{The formula could only be valid for $\ell=0$ if, for some reason, the growth of the four-point function of the $(A_1,A_2)$ theory, in the limit relevant for the dropping of arcs of integration along the derivation of the inversion formula, was better than the generic growth expected in any CFT and derived in \cite{Caron-Huot:2017vep}.}

We point out that the only input was the leading $t-$ and $u-$channels contributions and thus the resulting OPE coefficients are an approximation for sufficient large spin. Indeed, the neglected contribution of the long multiplets should behave like $\ell^{-\tau}$ for large spin, with $\tau \sim 2.4$, while the stress tensor contributes as $\tau=2$. Nevertheless, we see in figure \ref{Fig:ClCH} that starting from $\ell=4$ the analytical result is already inside the numerically allowed range for the OPE coefficient. This is shown clearly in figure \ref{Fig:Cl4} where the result of \eqref{eq:nonchiral_lead} for $\ell=4$ is shown as a dashed blue line, together with the numerically allowed range. For $\ell=2$, however, the result of \eqref{eq:nonchiral_lead} (blue dashed line in figure \ref{Fig:Cl2}) is clearly insufficient, as it is outside the numerically allowed region.
Note that the numerical results are not optimal yet, \ie, while they provide true bounds they have not yet converged, and the optimal bounds will be more restrictive. Thus, the fact that the $\ell=4$ estimate was inside the numerical bounds should not be taken to mean the subleading contributions are negligible for such a low spin. What is in fact surprising is that the estimates from \eqref{eq:nonchiral_lead} are so close to the numerically obtained ranges for such low values of the spin. These results leave us optimistic that better estimates can be obtained by providing a few of the subleading contributions, as was done in \cite{Simmons-Duffin:2016wlq} for the $3d$ Ising model.
The computation used to obtain figure \ref{Fig:ClCH} could be easily extended to obtain estimates for the OPE coefficients of the 
$\CC_{\frac12, 2r_0- \frac32 (\frac{\ell}2-\frac12,\frac{\ell}2)}$  multiplets, and also the dimensions and OPE coefficients of the remaining operators in \eqref{eq:selruleschiral}. One particularly interesting multiplet would be $\BB_{1, \frac75 (0,0)}$ since, as discussed before, we expect it to be absent in the $(A_1,A_2)$ theory. However, this corresponds to a spin zero contribution and thus convergence of the inversion formula is not guaranteed.

\subsection{Inverting the non-chiral OPE}
\label{sec:nonchiral_inv}

Next we turn to the non-chiral channel, where we have a decomposition in superconformal blocks, and so we must obtain a supersymmetric version of the inversion formula of \cite{Caron-Huot:2017vep}.
We consider the inversion of the $s-$channel OPE of \eqref{eq:nonchiral}, with the superblocks given by \eqref{eq:superblock},
\be 
\langle \phi(x_1) \bar{\phi}(x_2)  \phi(x_3) \bar{\phi}(x_4) \rangle = \frac{(z \zb)^{-\frac{\NN}{2}}}{x_{12}^{2 \Delta_\phi} x_{34}^{2 \Delta_\phi}} \left(\sum\limits_{\OO_{\Delta,\ell}} |\lambda_{\phi \bar{\phi} \OO_{\Delta,\ell}}|^2 g_{\Delta+ \NN,\ell}^{\NN,\NN}(z, \zb) \right)\,,
\label{eq:nonchiralforCH}
\ee 
where we are interested in taking $\NN=2$, but the same equation is also valid for $\NN=1$, and so all that follows generalizes easily to that case. Fortunately, the fact that, up to the overall prefactor $(z \zb)^{-\NN/2}$ in \eqref{eq:nonchiralforCH}, the blocks relevant for the $s-$channel decomposition are identical to bosonic blocks of operators with unequal dimensions makes the task of obtaining an inversion formula very easy.
We can use the results of \cite{Caron-Huot:2017vep} with small modifications: The Lorentzian inversion formula applies to the term between brackets in \eqref{eq:nonchiralforCH}, and the fact that the pre-factor is not the correct one for operators of unequal dimension plays a small role in the derivation of \cite{Caron-Huot:2017vep}. The only time the prefactor is considered is when bounding the growth of the correlator, needed to show the inversion formula is valid for spin greater than one. The modified prefactor here seems to ameliorate the growth: we are inverting $(z \zb)^{\tfrac{\NN}{2}}$ times a CFT correlator whose growth is bounded as discussed in \cite{Caron-Huot:2017vep}.
The condition $\ell > 1$ on the inversion formula \eqref{eq:CHgeneratingbos} came from the need to have $\ell$ large such that one could drop the arcs at infinity during the derivation of \cite{Caron-Huot:2017vep}. The prefactor's behavior in this limit means the inversion formula will be valid for all $\ell > 1 - \NN$, and the results we obtain for $\NN=2$ should be valid for all spins.
Apart from this, the prefactor will only play a role when representing the correlator by its $t-$ and $u-$channel OPEs.
As such we apply \eqref{eq:CHgeneratingbos} with 
\be 
\GG(z,\zb)= \sum\limits_{\OO_{\Delta,\ell}} |\lambda_{\phi \bar{\phi} \OO_{\Delta,\ell}}|^2 g_{\Delta+ \NN,\ell}^{\NN,\NN}(z, \zb) \,.
\ee
The $t-$ and $u-$channels of the correlator \eqref{eq:nonchiralforCH} are given by a non-chiral and chiral OPE respectively.
Using the crossing equation \eqref{eq:nonchiralnonchiral} we see that the $t-$channel expansion of $\GG(z,\zb)$ is
\be 
\GG(z,\zb) =  (z \zb)^{\frac{\NN}{2}} \left( \frac{z \zb }{(1-z)(1-\zb ) } \right)^{r_0} \sum\limits_{\Delta,\ell} |\lambda_{\phi \bar{\phi} \OO_{\Delta,\ell}}|^2  \GG_{\Delta,\ell}(1-z,1-\zb)\,,
\label{eq:Gtchan_nonchiral}
\ee
with the superblock given by \eqref{eq:superblock}. While the $u-$channel is given by
\be 
\GG(z,\zb) = (z \zb)^{r_0+\frac{\NN}{2}} \sum\limits_{\Delta,\ell} |\lambda_{\phi \phi \OO}|^2 g_{\Delta, \ell}\left(\frac{1}{z},\frac{1}{\zb}\right)\,.
\label{eq:Guchan_nonchiral}
\ee

Once again, the leading contributions to the $s-$channel spectrum at large spin, \ie, the leading contributions for  $\zb \to 1$ in \eqref{eq:Gtchan_nonchiral}, are from the  $t-$channel identity and stress-tensor multiplet. The subleading contributions in the $t-$channel come from long multiplets with $\Delta > \ell + 2$. On the other hand, the leading twist contribution in the $u-$ channel arises from the $\EE_{2r_0}$ and $\mathcal{C}_{0,2 r_0-1,\left(\frac{\ell }{2}-1,\frac{\ell }{2}\right)}$ multiplets, whose twists are all exactly $2r_0$, and so one should consider the infinite sum over $\ell$. 
From a lightcone computation, \eg, \cite{Li:2015rfa}, we expect an individual chiral operator of twist $\tau_c$ to contribute to the anomalous dimensions of the non-chiral operators at large $\ell$ as $\frac{(-1)^\ell}{\ell^{\tau_c}}$. Similarly, a non-chiral operator of twist  $\tau$ contributes to the same anomalous dimension at large $\ell$ as $\frac{1}{\ell^{\tau}}$. In the case at hand, $\tau=2$ for the stress-tensor multiplet and $\tau_c=2.4$ for each of the infinite number of leading operators in the chiral channel. The contribution of an individual chiral operator in the $u-$channel is thus subleading for sufficiently large spin. This is similar to what happened in section \ref{sec:chiral_inv}, and while in this case the dimensions of the operators are protected, their OPE coefficients are not. Indeed, the value of these OPE coefficients remains elusive, and the best estimate we have to go on comes from the numerically obtained bounds for the operators with $\ell \leqslant 20$ presented in figure \ref{Fig:ClCH}. An interesting possibility would be to attempt to combine the numerical ranges for low spin with the estimate for the large spin OPE coefficients obtained from \eqref{eq:nonchiral_lead}. The numerical bounds on the OPE coefficients would turn into an estimate, in the form of an interval, for the anomalous dimension; we leave this exploration for future work. Here we apply the inversion formula \eqref{eq:CHgeneratingbos} only to the exchange of the identity and stress-tensor multiplets
\be 
C^t(z,\beta) \supset \int\limits_0^1 \frac{\d \zb}{\zb^2} \kappa_{\beta} k_\beta^{1,1}(\zb) \dDisc\left[  \frac{(z \zb)^{r_0+1} }{((1-z)(1-\zb ))^{r_0} } \left( 1+ |\lambda_{\phi \bar{\phi} \hat{\CC}_{0,0}}|^2 \GG_{2,0}(1-z,1-\zb)\right)\right]\,,
\label{eq:CHgeneratingsuper}
\ee
where one should recall that $\Delta_{12}=\Delta_{34}= \frac{\NN}{2}=1$ when taking the double-discontinuity.

Like before, the exchange of the identity in \eqref{eq:CHgeneratingsuper} gives the existence of double-twist operators $\left[\phi \bar{\phi}\right]_{m,\ell}$, with dimensions 
\be
\Delta_{\left[\phi \bar{\phi}\right]_{m,\ell}} \underset{\ell \gg 1}{\longrightarrow} 2 r_0+2m + \ell \,.
\label{eq:nonchiraldoubletwist}
\ee
Computing the OPE coefficients from the identity exchange we find, for the leading twist operators,
\be
|\lambda_{\phi \bar{\phi} \left[\phi \bar{\phi}\right]_{0,\ell}}|^2 \underset{\ell \gg 1}{\longrightarrow} \frac{4^{2-\ell } r_0  (r_0)_{\ell -2} (2 r_0+1)_{\ell -2}}{(1)_{\ell -2} (r_0+\ell -2) \left(r_0+\frac{1}{2}\right)_{\ell -2}} \,,
\label{eq:OPEdoubletwistsuper}
\ee
which are precisely the OPE coefficients of generalized free field theory, now decomposed in superblocks instead of bosonic blocks.

\begin{figure}[htb!]
             \begin{center}           
              \includegraphics[scale=0.35]{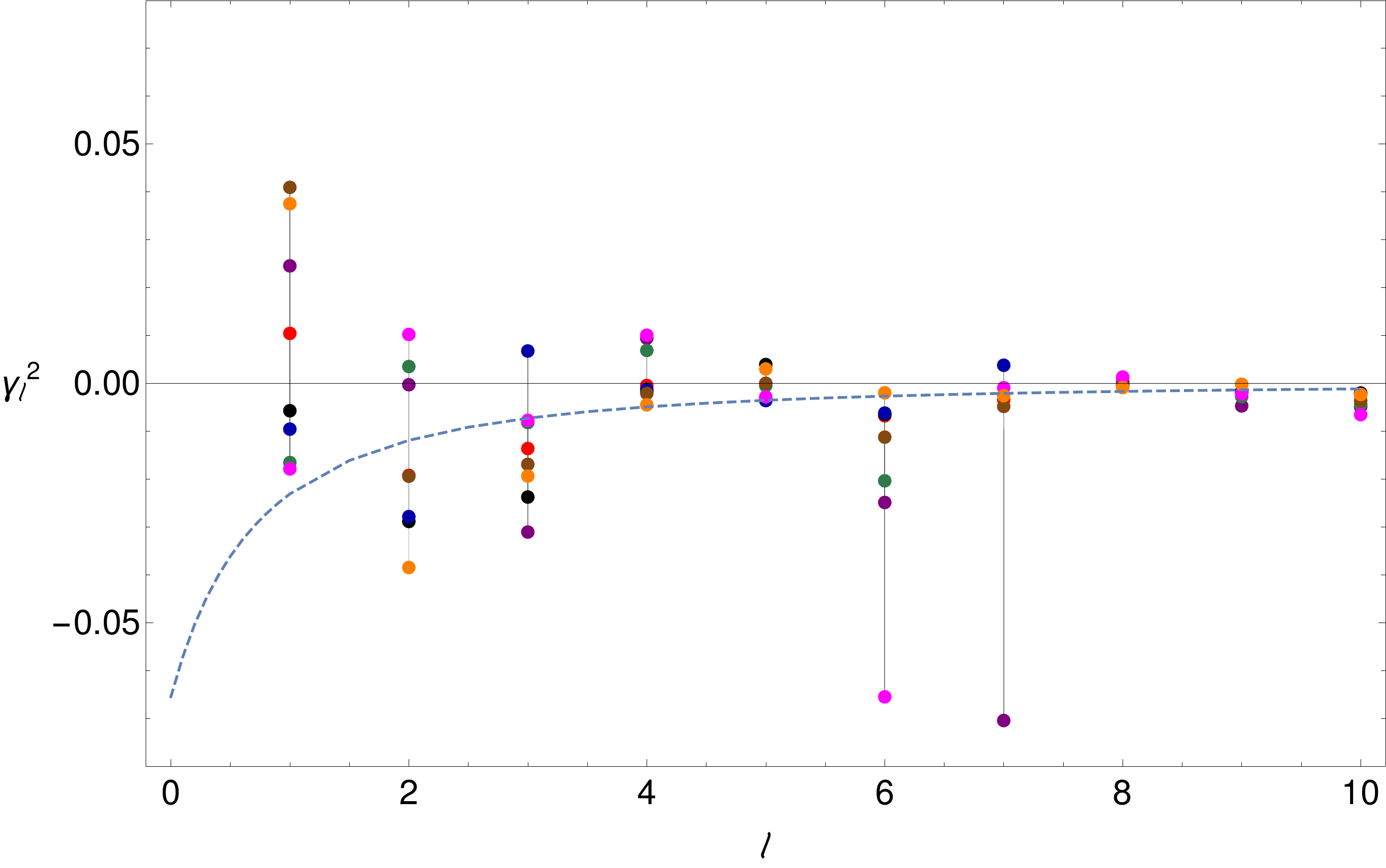}
              \caption{Anomalous dimension ($\gamma_\ell = \Delta_\ell - (2 \Delta_\phi + \ell)$) of the first spin $\ell$ long multiplet in the non-chiral channel. The colored dots are the dimension estimates extracted from the extremal functionals of the various bounds (figures \ref{Fig:c_min} and \ref{Fig:Bbound}-\ref{Fig:Cl24} for $c=\tfrac{11}{30}$) as indicated by their colors, with $\Lambda=34$. The dashed line corresponds to the result from the inversion formula \eqref{eq:CHgeneratingsuper}, for $c=\tfrac{11}{30}$ and external dimension $r_0=\frac65$, taking into account only the exchange of the identity and stress tensor in the $t-$channel, and is thus an approximate result for sufficiently large spin.
              }
              \label{Fig:anomCH}
            \end{center}
\end{figure}

The stress-tensor exchange provides corrections to these dimensions and OPE coefficients. As an illustration we computed its contribution to the anomalous dimensions of the leading twist operators $\left[\phi \bar{\phi}\right]_{0,\ell}$, $\gamma_\ell= \Delta_\ell - (2 \Delta_\phi + \ell)$. From the numerical estimates (see figure \ref{Fig:anomCH}) we see the anomalous dimensions starting at spin one are rather small, and so we simply take the zeroth order of the procedure outlined in \cite{Caron-Huot:2017vep} to commute the $z \to 0$ limit with the sum over primaries in \eqref{eq:CHgeneratingsuper}. These results are also shown in figure \ref{Fig:anomCH} for $\ell \geqslant1$ as a dashed blue line, together with estimates for these values arising from the various extremal functionals of section \ref{sec:numericsh0}, color coded according to which bound they came from.\footnote{We omitted two spin seven dimensions, as we could not accurately estimate them from the functionals. The two points that appear to be outlying in spin $6$ and $7$ correspond to cases where there were two zeros of the functional very close to one another, and we extracted the dimension of the first. We expect that higher derivative orders would fix both situations.} We are plotting the results starting from spin $\ell=1$. 
The leading $\ell=0$ operator is the stress tensor itself, which was not present in the generalized free field theory solution. As such the dimension of $\left[\phi \bar{\phi}\right]_{0,0}$ must come down from $2r_0=2.4$ to exactly $2$. The value of the anomalous dimensions coming from \eqref{eq:CHgeneratingsuper} is still insufficient for  this to happen, as clear from figure \ref{Fig:anomCH}.
For $\ell \geqslant1$, however, the numerical estimates of leading twist operators' dimensions are very close the values of double-twist operators \eqref{eq:nonchiraldoubletwist}. Indeed, the maximum anomalous dimension in figure \ref{Fig:anomCH}, ignoring the two out-lying points, is of the order of $\gamma_1 \sim 0.04$, in a dimension that is close to $2r_0+1 =3.4$.
The anomalous dimensions obtained from \eqref{eq:CHgeneratingsuper} (dashed blue line in \eqref{Fig:anomCH}) are close to the numerically obtained values starting from  $\ell=2$, despite the fact that our results are only valid for sufficiently large spin, as we have only considered the identity and stress tensor contributions in the $t-$channel, and completely disregarded any $u-$channel contribution.  In particular, for spin $\ell \gtrsim 8$ the numerical estimates arising from the different extremization problems of section \ref{sec:numericsh0} are all cluttered, approaching the value \eqref{eq:nonchiraldoubletwist}, and close to the values coming from \eqref{eq:CHgeneratingsuper}.

\bigskip

All in all, we have seen that both in the chiral and non-chiral channels the estimates coming from applying the inversion formula, and providing only the leading twist operators (identity plus stress-tensor supermultiplet), come very close to the numerically obtained bounds/estimates. This leaves us optimistic that the spectrum of the $(A_1,A_2)$ can be bootstrapped, similarly to the $3d$ Ising model. The numerical results for $\NN=2$ theories suffer from slow convergence and thus the estimates for OPE coefficients and anomalous dimensions we obtain are not yet with the precision of those of the $3d$ Ising model. By using this data as input to the inversion formulas they would in turn produce ranges for the various quantities appearing in the chiral and non-chiral OPEs.
Finally, another direction corresponds to using the output of each inversion formula as input for the other to obtain better estimates. We leave these two directions for future work.

\acknowledgments

We have greatly benefited from discussions with
Philip Argyres,
Lorenzo Bianchi,
Mario Martone,
Marco Meineri,
and 
Volker Schomerus.
The authors thank the organizers of the Pollica Summer Workshop 2017 for hospitality, and were partly supported by the ERC STG grant 306260 during the Pollica Summer Workshop.
M.L. and P.L. thank ICTP-SAIFR in S\~{a}o  Paulo for hospitality during the Bootstrap 2017 workshop, and the Simons Collaboration on the Non-perturbative Bootstrap for providing many stimulating workshops and conferences during which this work was carried out. We also thank the participants of the Simons Collaboration's kick off, ``Exact Operator Algebras in Superconformal Field Theories'' and ``Numerical bootstrap'' meetings for discussions. M.L. was partially supported  by the German Research Foundation (DFG) via the Emmy Noether program ``Exact results in Gauge theories''.

\appendix
\section{Blocks and Crossing}
\label{sec:blocksandcrossing}

We write the bosonic blocks for the exchange of a conformal primary of dimension $\Delta$ and spin $\ell$, in the four-point function of unequal scalar operators of dimensions $\Delta_{i=1,\dots,4}$, as \cite{Dolan:2000ut}
\bea
g_{\Delta,\ell}^{\Delta_{12},\Delta_{34}}(z,\zb) &=& \frac{z \bar{z}}{z-\bar{z}}\left( k^{\Delta_{12},\Delta_{34}}_{\Delta+\ell}(z) k^{\Delta_{12},\Delta_{34}}_{\Delta-\ell-2}(\bar{z})-z \leftrightarrow \bar{z} \right)\,,\nn\\
k^{\Delta_{12},\Delta_{34}}_{\beta}(z) &=& z^{\frac{\beta}{2}} {_2F_1}\left(\frac{\beta -\Delta_{12}}{2},\frac{\beta +\Delta_{34}}{2};\beta ;z\right)\,,
\label{eq:bosblock}
\eea
where  $\Delta_{ij}=\Delta_i -\Delta_j$, and $z$ and $\zb$ are obtained from the standard conformally invariant cross-ratios 
\be 
z \zb = \frac{x_{12}^2 x_{34}^2}{x_{13}^2 x_{24}^2}\,, \qquad \qquad 
(1-z)(1-\zb) = \frac{x_{14}^2 x_{23}^2}{x_{13}^2 x_{24}^2} \,.
\ee

The crossing equations \eqref{eq:crossingeqs} (see \cite{Beem:2014zpa,Lemos:2015awa} for a derivation) are written here in a form suitable for the numerical analysis of section \ref{sec:numericsh0}
\be
\sum_{\OO \in \phi \bar \phi} | \lambda_{\phi \bar \phi \OO} |^2
\begin{bmatrix}
(-1)^\ell\tilde{\FF}_{\pm, \Delta,\ell}(z,\zb)\\
 \FF_{-, \Delta,\ell}(z,\zb)
\end{bmatrix}
+
\sum_{\OO \in \phi \phi} |\lambda_{\phi \phi \OO} |^2
\begin{bmatrix}
\mp (-1)^\ell F_{\pm, \Delta,\ell}(z,\zb)\\
0
\end{bmatrix}
 = 0\,,
\label{eq:cross_numerics}
\ee
where the first line encodes two separate crossing equations, differing by the signs indicated, and where we defined (recall that $\Delta_\phi=\Delta_{\bar{\phi}}=r_0$)
\be
\begin{split}
F_{\pm, \Delta,\ell}(z,\zb) &\equiv \left((1-z)(1-\zb)\right)^{r_0} g^{0,0}_{\Delta,\ell}(z,\zb) \pm (z \zb)^{r_0} g^{0,0}_{\Delta,\ell}(1-z,1-\zb)\,,\\
\FF_{\pm, \Delta,\ell}(z,\zb) &\equiv \left((1-z)(1-\zb)\right)^{r_0} \GG_{\Delta,\ell}(z,\zb) \pm (z \zb)^{r_0} \GG_{\Delta,\ell}(1-z,1-\zb)\,,\\
\tilde{\FF}_{\pm, \Delta,\ell}(z,\zb) &\equiv \left((1-z)(1-\zb)\right)^{r_0} \tilde{\GG}_{\Delta,\ell}(z,\zb) \pm (z \zb)^{r_0}  \tilde{\GG}_{\Delta,\ell}(1-z,1-\zb)\,,
\end{split}
\ee
with the superblocks $\GG_{\Delta,\ell}$ and $\tilde{\GG}_{\Delta,\ell}$ given in  \eqref{eq:superblock} and \eqref{eq:superblockbraid}.
In \eqref{eq:cross_numerics} the stress tensor and the identity contribute as
\be
\overrightarrow{V}_{\mathrm{fixed}} = 
\begin{bmatrix}
\tilde \FF_{\pm , \Delta=0,\ell=0}(z,\zb)\\
\FF_{- , \Delta=0,\ell=0}(z,\zb)
\end{bmatrix} +
\frac{r_0^2}{6c}
\begin{bmatrix}
\tilde \FF_{\pm , \Delta=2,\ell=0}(z,\zb)\\
 \FF_{- , \Delta=2,\ell=0}(z,\zb)
\end{bmatrix}\,.
\label{eq:idandst}
\ee

\bibliography{./aux/biblio}
\bibliographystyle{./aux/JHEP}

\end{document}